\documentclass[twocolumn,twocolappendix]{aastex701}

\newcommand{\uu}{\mbox{\boldmath $u$} {}}
\begin{document}

\title{On the Relationship between Solar Spicules and Propagating Coronal Disturbances: The Role of Shocks}

\author[orcid=0009-0002-0549-1446,gname=Ravi,sname=Chaurasiya]{Ravi Chaurasiya}
\altaffiliation{Both authors contributed equally to this work}
\affiliation{Udaipur Solar Observatory, Physical Research Laboratory, Udaipur-313001, India}
\affiliation{Indian Institute of Technology, Gandhinagar, Gujarat-382355, India}
\email{ravi@prl.res.in}

\author[orcid=0009-0000-2614-254X,gname=Sankalp,sname=Srivastava]{Sankalp Srivastava}
\altaffiliation{Both authors contributed equally to this work}
\affiliation{Indian Institute of Astrophysics, $2^{nd}$ Block, Koramangala, Bengaluru-560034, India}
\affiliation{Pondicherry University, R.V.Nagar, Kalapet, Puducherry-605014, India}
\email{sankalp.srivastava@iiap.res.in}

\author[orcid=0000-0002-0181-2495,gname=Piyali,sname=Chatterjee]{Piyali Chatterjee}
\affiliation{Indian Institute of Astrophysics, $2^{nd}$ Block, Koramangala, Bengaluru-560034, India}
\affiliation{Pondicherry University, R.V.Nagar, Kalapet, Puducherry-605014, India}
\email{piyali.chatterjee@iiap.res.in}
\author[orcid=0000-0002-3369-8471,gname=Sahel,sname=Dey]{Sahel Dey}
\affiliation{School of Science, University of Newcastle, University Drive, Callaghan, NSW 2308, Australia}
\email{sahel.dey@newcastle.edu.au}

\author[orcid=0000-0003-3439-4127,gname=Robertus,sname=Erd\'elyi]{Robertus Erd\'elyi}
\affiliation{Solar Physics and Space Plasma Research Centre (SP2RC),
School of Mathematical and Physical Sciences,
University of Sheffield, Sheffield S3 7RH, UK}
\affiliation{Department of Astronomy, Eötvös Loránd University,
Budapest, Pázmány P. sétány 1/A, H-1117, Hungary}
\affiliation{Gyula Bay Zoltán Solar Observatory (GSO),
Hungarian Solar Physics Foundation (HSPF),
Petőfi tér 3., Gyula H-5700, Hungary}
\email{r.von.fay-siebenburgen@sheffield.ac.uk}

\author[orcid=0000-0001-5802-7677,gname=Ankala,sname=Raja Bayanna]{Ankala Raja Bayanna}
\affiliation{Udaipur Solar Observatory, Physical Research Laboratory, Udaipur-313001, India}
\email{bayanna@prl.res.in}

\begin{abstract}

Spicules and propagating coronal disturbances (PCDs) are ubiquitous dynamic features of the solar atmosphere, yet their physical connection remains an open question of paramount importance to the mass and energy transport in the solar atmosphere. Using concurrent multiwavelength high-resolution observations from the Swedish 1-m Solar Telescope and the Solar Dynamics Observatory, supported with two-dimensional radiative magnetohydrodynamic (MHD) simulations, we find that i) shock waves in the chromosphere generated from non-linear wave steepening drive some spicules, ii) in the corona, these shock waves may transition into large amplitude non-linear compressive MHD waves depending on the magnetic field strength and the ambient coronal conditions. In either case, the shocks or the large-amplitude compressive waves in the corona, also transport upward mass flux and produce intensity variations in the form of PCDs in coronal passbands. Further a multi-height wavelet analysis shows dominant $\sim$5 minute periods in the lower chromosphere that evolve into longer periods ($\ge$10 minutes) at higher atmospheric layers, consistent with dispersive propagation in a stratified medium. The observational characteristics together with the numerical simulations, demonstrate that a shock-driven MHD mechanism links spicule formation to coronal disturbances. Finally, mass flux estimates from both the observations and the simulations indicate that these PCDs can also aid in supplying mass to the solar wind.
\end{abstract}


\section{Introduction} \label{sec:introduction}
The solar atmosphere is a highly dynamic environment where magnetic fields and plasma flows interact in complex ways, giving rise to a variety of fine-scale structures. Among these spicules, which are thin jet-like features observed prominently at the solar limb \citep{Beckers72,Tsiropoula.et.al.12} and on the solar disk as mottles or dynamic fibrils \citep{Tsiropoula.et.al.94,Suematsu.et.al.95,Hansteen.et.al.06} are of particular interest due to their ubiquity and their potential role in mass and energy transport into the solar corona \citep{Carlsson.et.al.19,Ni26}. Spicules exhibit rapid upward and downward motions and, in many cases, are heated to transition region (TR) and coronal temperatures \citep{Pereira.et.al.14,RouppevanderVoort.et.al.15,Chaurasiya.et.al.24,Bose.et.al.25}. 

Although significant progress has been made in understanding the heating of spicular plasma to such high temperatures, the precise physical mechanisms responsible for their formation and dynamics remain a subject of debate \citep{Carlsson.et.al.19}. Several mechanisms have been proposed to explain the origin and driving of these spicules. These include shock waves generated by the leakage of photospheric oscillations into the chromosphere \citep{DePontieu.et.al.04,Hansteen.et.al.06, Heggland.et.al.2011, Gopalan.Priya.et.al.18, Dey.et.al.22}. Recent study by \citet{Sankalp.srivastava.et.al.25} found that the strength of the shock is well correlated with the height of the corresponding spicule in their two-dimensional radiative magnetohydrodynamic (rMHD) simulations. Alternative drivers of spicules could be Alfv\'en waves or pulses \citep{Ijima.etal.2017,Oxley.et.al.20,Sakaue&Shibata20,Srivastava.et.al.24}, and magnetic reconnection in small-scale magnetic configurations \citep{Ding.et.al.11,Yurchyshyn.et.al.13,Samanta.et.al.19,Gonzalez-Aviles2020}. Other processes such as granular buffeting \citep{Roberts79}, rebound shocks \citep{Hollweg82, Murawski.et.al.2010}, chromospheric swirls \citep{Liu.et.al.19}, the release of amplified magnetic tension facilitated by ambipolar diffusion \citep{Martinez.sykora.et.al.17}, and Ellerman bombs \citep{Sand.et.al.25} have also been suggested as possible drivers. Although each mechanism provides valuable insight, it is more likely that multiple processes operate simultaneously or under different solar conditions, giving rise to the dynamic nature of the spicules.

In addition to their importance in chromospheric \& TR dynamics, spicules are also believed to play a role in shaping disturbances in the solar corona. Observations have revealed upward-propagating disturbances in coronal structures, appearing as intensity perturbations, that have also been linked to spicular activity (e.g.,  \citealt{Jiao.et.al.15,Skirvin.et.al.24}). There is an indication that both waves and flows might be responsible for the origin of these propagating coronal disturbances (PCDs), however, the debate has not yet been fully settled (e.g., \citealt{Bryans.et.al.16,Martinez.sykora.et.al.2018, Banerjee.etal.2021}). 
Since the lower solar atmosphere is replete with shock waves, it can be expected that they possibly serve an important role in linking spicules and PCDs. Furthermore, recent quiet-Sun chromospheric modeling has suggested that a conversion of chromospheric acoustic shocks back into linear waves might introduce considerable slow MHD wave flux into the corona \citep{noraz2026A&A}. However, a comprehensive physical picture linking the spicules and spicule-driving shocks seen in the lower atmosphere to the PCDs is still missing, which can be crucial for understanding whether spicules play a role in supplying energy and mass to heat the solar corona and drive the solar wind \citep{Withbroe&Noyes.77}, the primary agent of space weather. 

In this work, we focus on studying the spicules driven by shock waves and their role in triggering PCDs. We investigate this connection among spicules, shocks and PCDs through high-resolution multi--wavelength observations and further support our findings with insights from two-dimensional rMHD simulations. We also present mass flux estimates from both observations and simulations and show how they relate to the mass loss rate of the solar wind.

\begin{figure}
\includegraphics[width=85mm]{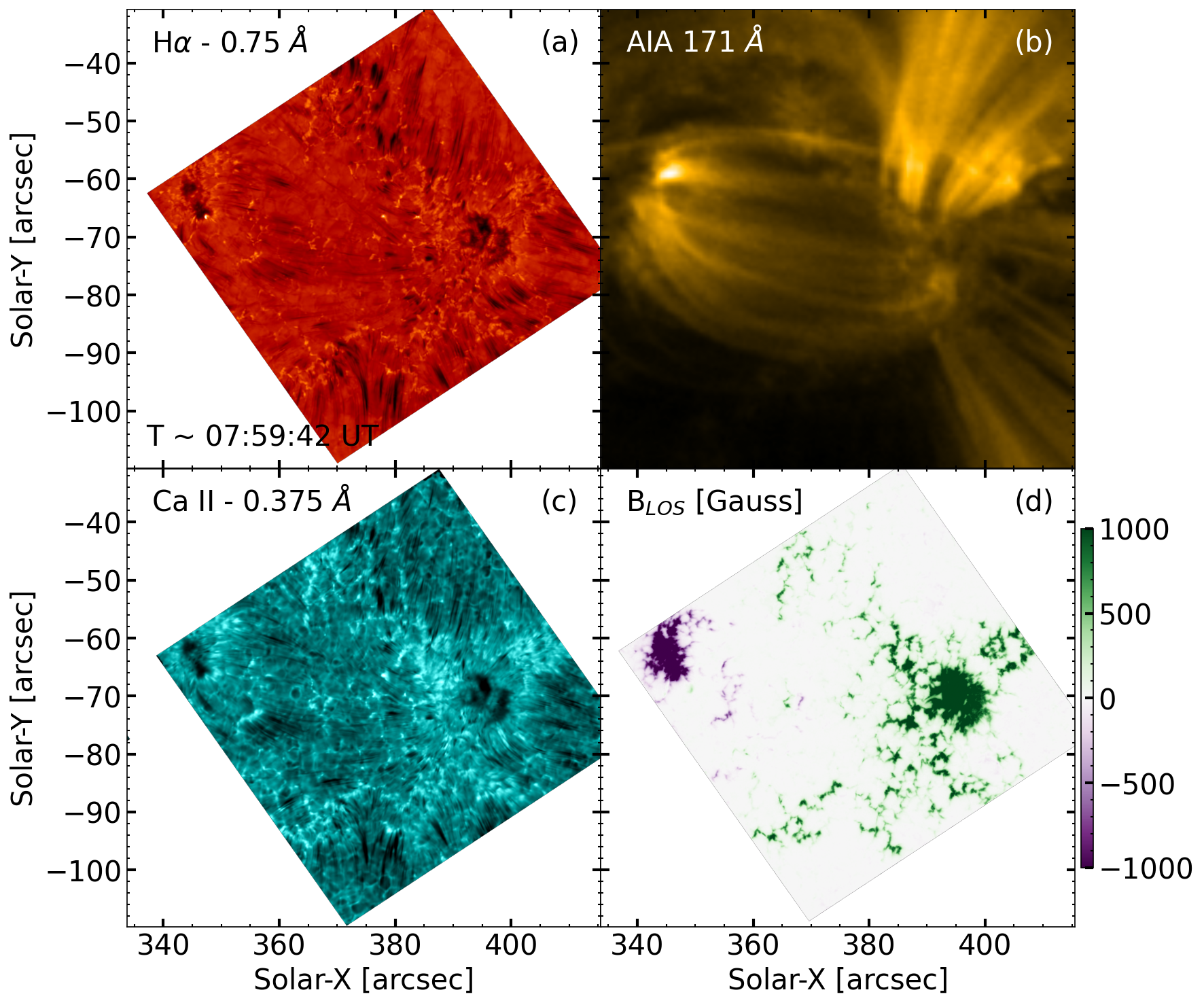}
\caption{Panels (a) and (c) show the chromosphere of the active region as observed in the blue wings of H$\alpha$ (H$\alpha$$-0.8$ \AA) and Ca \textsc{ii} IR (Ca \textsc{ii} IR$-0.375$ \AA), respectively. Panels (b) and (d) display the lower corona as observed in AIA~171 \AA\ and the photospheric magnetic field inferred from PyMilne inversion (\cite{delaCruzRodriguez.19}) of Fe \textsc{i}~6173 \AA, respectively. The spicules may be identified as elongated absorption features appearing as dark threads over the Panels (a) and (c).}
\label{Fig:Data_overview}
\end{figure}

\section{Association of Shock-Driven Spicules with PCDs: observational signatures}\label{sec:observation}

We analyzed a coordinated dataset of NOAA Active Region 12760, observed on 30 April 2020 from 07:58 UT to 09:37 UT using the Swedish 1-m Solar Telescope (SST; \citep{Scharmer.et.al.03}) and the Solar Dynamics Observatory (SDO; \citep{Pesnell.et.al.12}). The heliocentric coordinates of the target region were approximately ($374\arcsec$, $-70\arcsec$), corresponding to an observing angle of $\mu = 0.92$. The active region contained multiple solar pores and network structures with a moderate magnetic field strength (see Figure~\ref{Fig:Data_overview}).
The field of view of the observations is approximately $60\arcsec \times 60\arcsec$. Further details of the SST and SDO observations, including spectral sampling, cadence, alignment procedures and the inversion, are provided in Appendix~\ref{app:obs}.

Motivated by earlier studies identifying sawtooth patterns in $\lambda$–$t$ diagrams as a characteristic observational signature of shock waves in the lower solar atmosphere \citep{Carlsson&Stein97,RouppevanderVoort.et.al.03,Chaurasiya.et.al.25}, we explore the role of shock waves in driving spicules and examine their relation with propagating coronal disturbances (PCDs) within the region under study.

\begin{figure*}
\centering
\includegraphics[width=40mm]{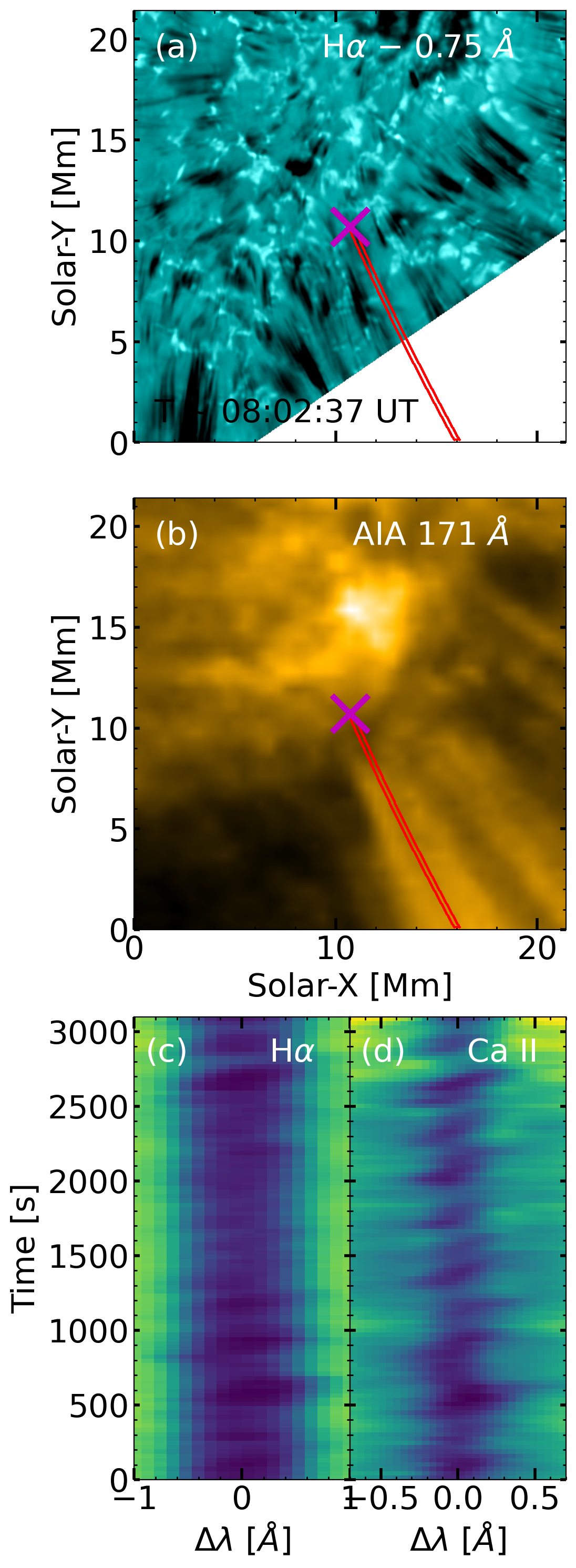}
\includegraphics[width=68mm]{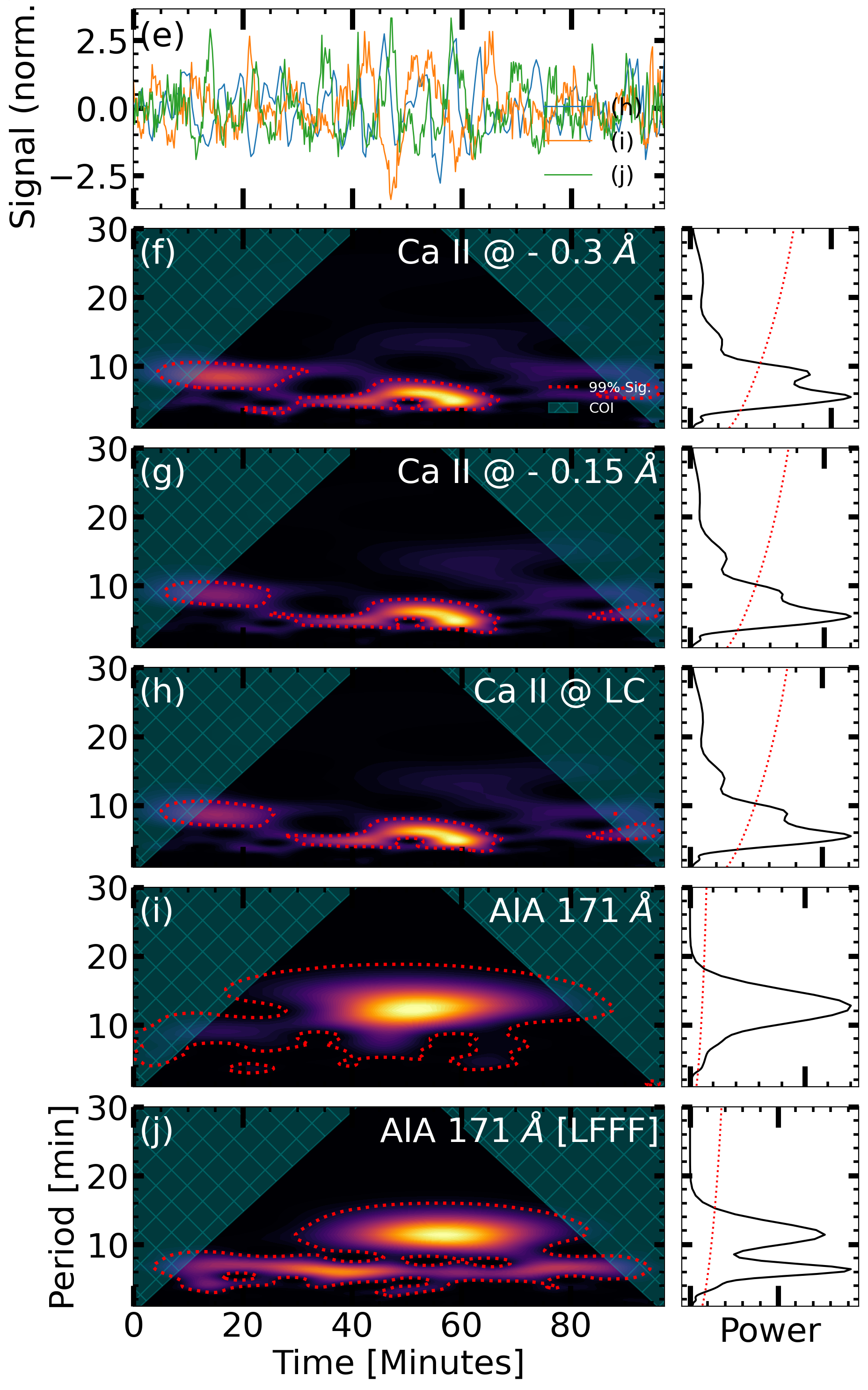}
\caption{Panels (a) and (b) show the chromospheric and the low coronal structures of the solar atmosphere as observed in the blue wing of H$\alpha$ and AIA 171~\AA, respectively. Panels (c) and (d) present the $\lambda$--$t$ diagrams of the H$\alpha$ and Ca\,\textsc{ii}~8542~\AA\ spectral lines at the location marked by the cross (X) in Panel (a). The red slit (or box) in Panels (a) and (b) indicates the region used to construct the $x$--$t$ map shown in Fig.~\ref{Fig:Shock_PCD_ex1_space_time}. Panel (e) shows the time series of the normalized Doppler signals of the Ca\,\textsc{II}~8542~\AA\ LC and AIA 171~\AA~ light curves only for clarity. Panels (f)--(h) display the wavelet analysis of the Ca\,\textsc{ii}~8542~\AA\ Doppler velocity timeseries (indicated in panel (e)), obtained at multiple heights at the shock-wave locations covering lower to upper chromosphere, whereas Panel (i) presents the wavelet analysis of the corresponding AIA 171~\AA~light curve (shown in panel (e)). Panel (j) also presents the wavelet analysis of the AIA 171~\AA~ light curve, but for the location obtained from the LFFF extrapolation, as indicated by the red magnetic field line in Figure~\ref{Fig:LFF}.}
\label{Fig:Merge_fig2}
\end{figure*}

We first manually inspected the $\lambda$–$t$ diagrams at the footpoints of spicules to investigate whether the spicules originate from locations associated with chromospheric shock waves. Figure~\ref{Fig:Merge_fig2} shows concurrent observations of the low chromosphere and the low corona in the H$\alpha$ $-0.75$~\AA\ wing and the AIA 171~\AA\ channel, together with the $\lambda$–$t$ diagrams in H$\alpha$ and Ca~\textsc{ii} 8542~\AA\ corresponding to the base of the slit (marked by the magenta X). From panel (a), it is evident that the footpoint of a spicule spatially coincides with a location exhibiting shock-wave signatures, characterized by episodic sawtooth patterns in time. It appears that the characteristic sawtooth patterns are more prominent in the Ca~\textsc{II} 8542~\AA\ data primarily because the Ca~\textsc{II} 8542~\AA\ line has a narrower core compared to the broader H$\alpha$ line core, making shock signatures easier to detect. In addition, the Ca~\textsc{II} 8542~\AA\ observations have a finer spectral sampling ($\sim$7.5 m\AA) compared to the H$\alpha$ observations ($\sim$25 m\AA). The combined effect of the broader line core and coarser spectral sampling in H$\alpha$ reduces the ability to resolve rapid spectral variations associated with shock propagation. As a result, the shock-wave signatures (characteristic sawtooth patterns) appear much clearer in the Ca~\textsc{II} 8542~\AA\ $\lambda$--$t$ diagrams than in the H$\alpha$ data. These sawtooth patterns indicate strong chromospheric upflows followed by subsequent downflows.

\begin{figure*}
\centering
\includegraphics[width=0.7\textwidth]{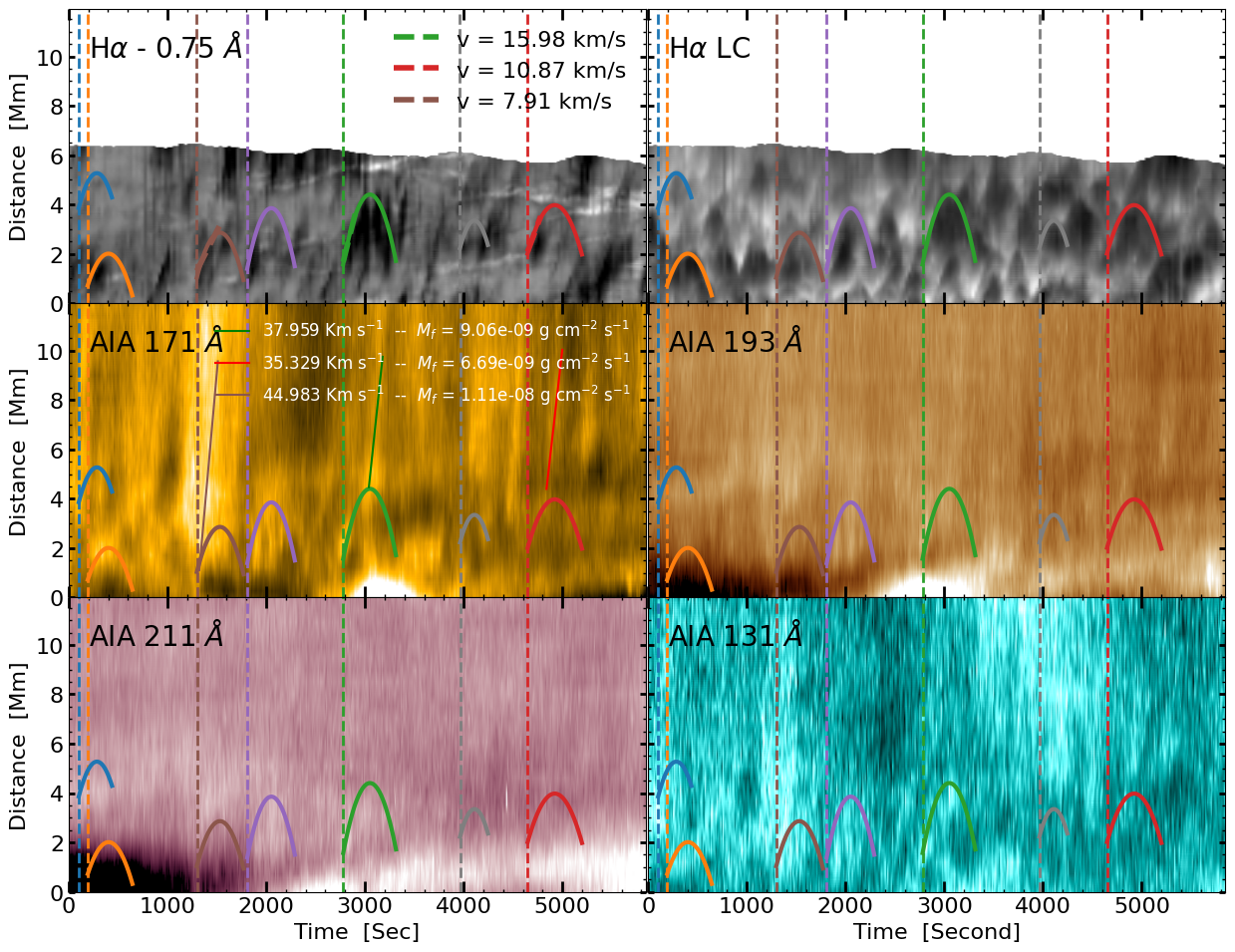}
\caption{Space–time maps obtained from the virtual slit (see Figure~\ref{Fig:Merge_fig2}) corresponding to the H$\alpha$ wing, H$\alpha$ line center and various SDO/AIA channels. The colored parabolas trace the parabolic trajectories of different spicules. The dashed line on the left side of each parabola is used to estimate the projected velocity. The vertical dashed colored lines mark the onset of shock driven spicular activity or the beginning of PCDs, indicating that the PCDs are triggered by the shock driven spicules.}
\label{Fig:Shock_PCD_ex1_space_time}
\end{figure*}

To further investigate the possible connection between shock-driven spicules in the 
chromosphere and PCDs in the corona, we constructed space–time ($x$–$t$) maps by placing a 
virtual slit of width 7 pixels (corresponding to approximately 0.3~Mm) across the chromosphere 
and the corresponding SDO/AIA channels, as shown in Figure~\ref{Fig:Shock_PCD_ex1_space_time}. The resulting $x$–$t$ maps of the H$\alpha$ line core and wing clearly illustrate that spicules trace parabolic trajectories as part of their evolution. The projected initial velocities of the spicules range between $\sim$ 7.91 and $\sim 15.98$\,km\,s$^{-1}$. The signatures of PCDs in the corona are identified in the $x$–$t$ diagrams of multiple AIA filters, revealing a temporal coincidence between the onset of the PCDs and shock-driven spicules. This correspondence indicates that shock-driven spicular activity plays a pivotal role in triggering PCDs. It is also evident that PCDs follow approximately a linear characteristic curve in $x$–$t$ maps, which refers to constant projected propagation speed between $\sim$ 37.95 and $\sim 44.98$\,km\,s$^{-1}$ for different events. Two additional examples similar to Figure~\ref{Fig:Merge_fig2} are presented in Figures~\ref{Fig6A} and~\ref{Fig7A}, with their corresponding space–time maps shown in Figures~\ref{Fig8A} and~\ref{Fig9A}, respectively.

The wavelet analysis of the multi-height Doppler velocities sampling different atmospheric layers in the chromosphere (panels e–g) and the AIA 171~\AA\ intensity signal sampling the lower corona (panel h; assuming vertical propagation), for this shock event is also presented in Figure~\ref{Fig:Merge_fig2}. The details of the estimation of the multi-height Doppler velocities are provided in Appendix~\ref{sec:Doppler}. However, since coronal magnetic field lines are generally inclined rather than strictly vertical, we additionally performed wavelet analysis of the AIA 171~\AA\ signal along the magnetic-field-aligned trajectory obtained from Linear Force-Free Field (LFFF; see Appendix Section~\ref{app:LFF} for details) extrapolation, as shown in panel (j). The extrapolated magnetic field geometry is illustrated in Figure~\ref{Fig:LFF}. Notably, the wavelet power distribution changes slightly when the signal is sampled along the inclined magnetic field line compared to the vertically projected location above the chromospheric pixels shown in panels (i).

\section{Insights from Radiative MHD Simulations}\label{sec:simulation}

The relationship among the spicules, shocks and PCDs is further explored utilizing a fully compressible 
radiative MHD model, which considers multiple magnetic network regions. The simulation setup (in {\sc Pencil 
Code}\footnote{\href{https://pencil-code.org/}{https://pencil-code.org/}}) is a 2D Cartesian box covering a 
vertical range from 5 Mm below the photosphere to 31 Mm above. The vertical extent includes the upper 
convection zone, photosphere, chromosphere, TR, and lower corona. The horizontal extent of the domain is 18 
Mm, which captures several convective granulation cells. The uniform grid spacing is 24 km in both horizontal and 
vertical direction. The model incorporates the essential physics needed to comprehensively model the 
highly stratified solar atmosphere, including Spitzer thermal conduction, an ionized equation of state (with ionization fraction calculated using 
Saha ionization formula), solving the detailed LTE radiative transfer equation for the optically thick regime, and tabulated radiative cooling in the optically thin upper atmosphere.

\begin{figure*}[ht!]  
\centering
\includegraphics[width=0.7\textwidth]{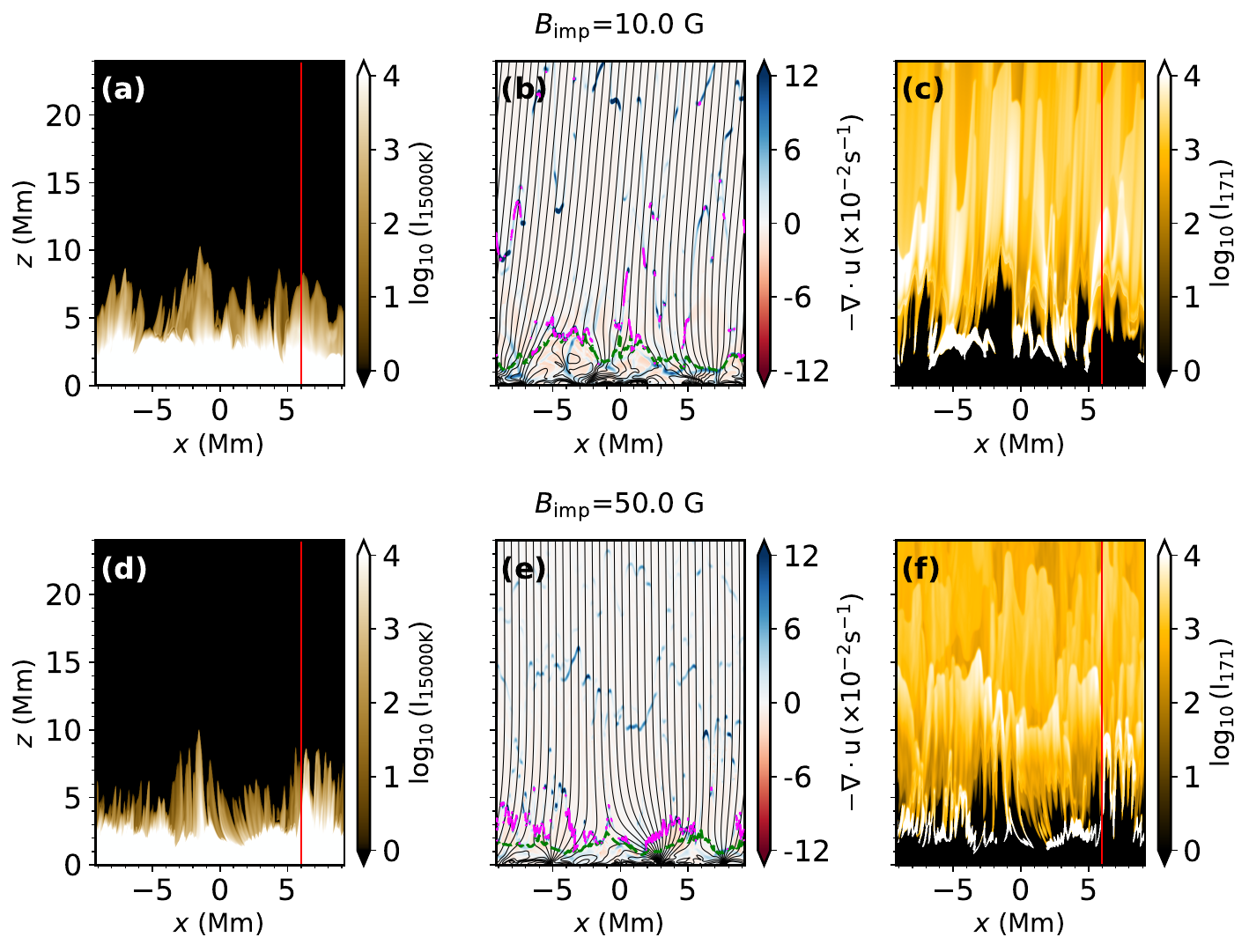}
    \caption{Panels (a)-(c) show a snapshot at a time $t=$83.3 minutes from the start of the
    simulation with an imposed vertical magnetic field $B_{\mathrm{imp}}$= 10.0 G. The 
    spicules are visible in the synthetic emission at 15,000 K in (a), while the associated 
    PCDs can be seen in the synthetic AIA 171 \AA~emission in (c). The slit used for obtaining 
    the time-distance diagrams in Figure \ref{fig:space-time:simulation} is depicted by red lines 
    in both (a) and (c). Panel (b) depicts the velocity convergence regions (blue). Additionally, 
    the strong convergence regions which qualify as slow shocks are over-plotted in magenta. The 
    plasma-$\beta$=1 contour is shown as a green dotted line. Panels (d)-(f) are similar to (a)-(c) 
    but for $B_{\mathrm{imp}}$= 50.0 G at $t=$83.3 minutes from start of simulation. (An animation of this 
    figure for 60 minutes of solar time is available in the online article.)}
    \label{fig:simulation_context}
\end{figure*}

The overall magnetic field resembles an open field topology, with an imposed vertical background field $B_\mathrm{imp}$. We analyze two 
different models with $B_\mathrm{imp}=10 \, \mathrm{G}$ and $B_\mathrm{imp}=50 \, \mathrm{G}$ while 
keeping all other parameters the same, and compare the results. The horizontal boundaries are periodic, while the top 
boundary is maintained at a constant temperature of 1 MK. A sponge layer in the uppermost 6 Mm of the domain is 
included to absorb any reflection of the outgoing waves from the top boundary. No additional magnetic flux 
emergence is employed in the domain. Further details regarding the model can be found in \cite{Dey.et.al.22}, \cite{kesri2024}, and \cite{Sankalp.srivastava.et.al.25}. An example of the relaxed 
temperature and density structuring in the simulation setup for the $B_\mathrm{imp}=10 \, \mathrm{G}$ case 
is shown in Figure \ref{fig:density-temp:simulation} of the Appendix, after the convection has saturated. The snapshot corresponds to the first snapshot of the analyzed time series. The model is able to self-consistently excite a forest of spicules as a result of the sub-surface convective processes, as shown in Figure \ref{fig:simulation_context}. We analyzed a 60 minutes long time series of snapshots from the simulations, taken at a cadence of 20 s, to study the link between the spicules and shocks and how this relationship manifests itself in the corona when observed through EUV synthetic emission. In Figure~\ref{fig:simulation_context}~(panels a and d), spicules are seen as enhanced fine-scale emission structures at 15,000 K for 10 G and 50 G imposed magnetic fields, respectively. The slow shock wave fronts, which propagate along the local magnetic field, are identified
in the simulation domain using the automated shock detection technique (discussed in Appendix~\ref{app:auto_shock}) and marked with magenta points (b and e panels of Figure~\ref{fig:simulation_context}). 
The leakage of global solar oscillations at the photosphere cascades into the highly stratified atmosphere above and steepens to shock waves. 
In our model, spicular jets are primarily driven and accelerated by these non-linear shock waves that are found to be located at the spicule tips during the spicule rise phase. These localized fronts are associated with strong flow convergence regions, as shown in panels (b) and (e) of Figure~\ref{fig:simulation_context} for 10 G and 50 G imposed magnetic fields, respectively. However, not all converging flow regions correspond to shock waves. By definition, at a slow shock front, the flow speed in the rest frame of the shock must jump from {\em superslow} (greater than slow MHD speed) to {\em subslow} (less than slow MHD speed) while moving from upstream to the downstream region.

For understanding the impact of the spicules and/or propagating shock waves in the corona, we investigate a simultaneous plasma emission map, similar to the coronal temperature-sensitive AIA 171 \AA ~channel (panels c and f of Figure~\ref{fig:simulation_context}). The coronal emission reveals the highly structured nature of the atmosphere in the presence of spicules and shock structures for both setups. Some of the enhanced emission features are possibly connected to PCDs and are detected much higher in the corona for the 10 G case. 

\begin{figure*}
\includegraphics[width=0.69\textwidth]{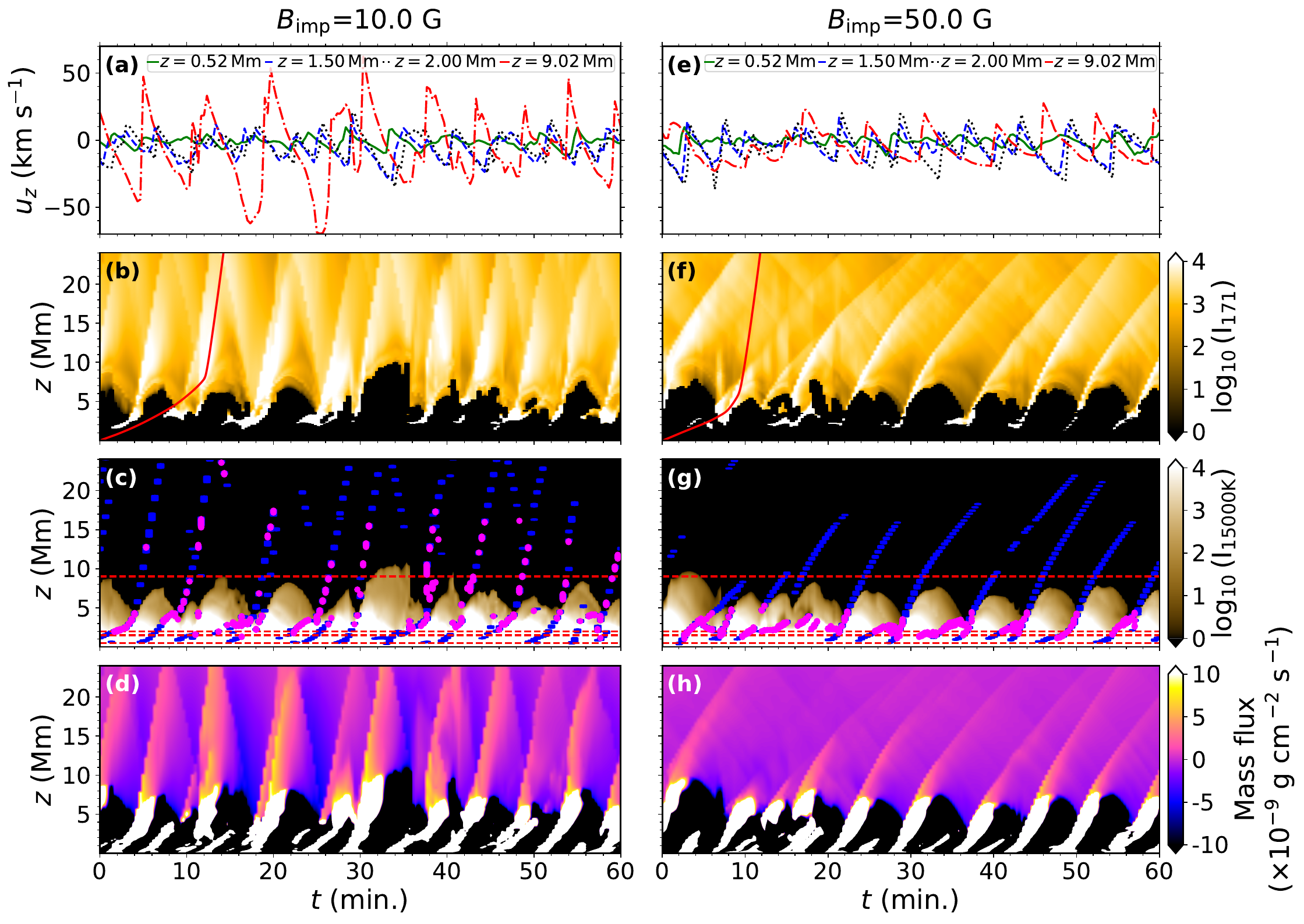} 
\includegraphics[width=0.31\textwidth]{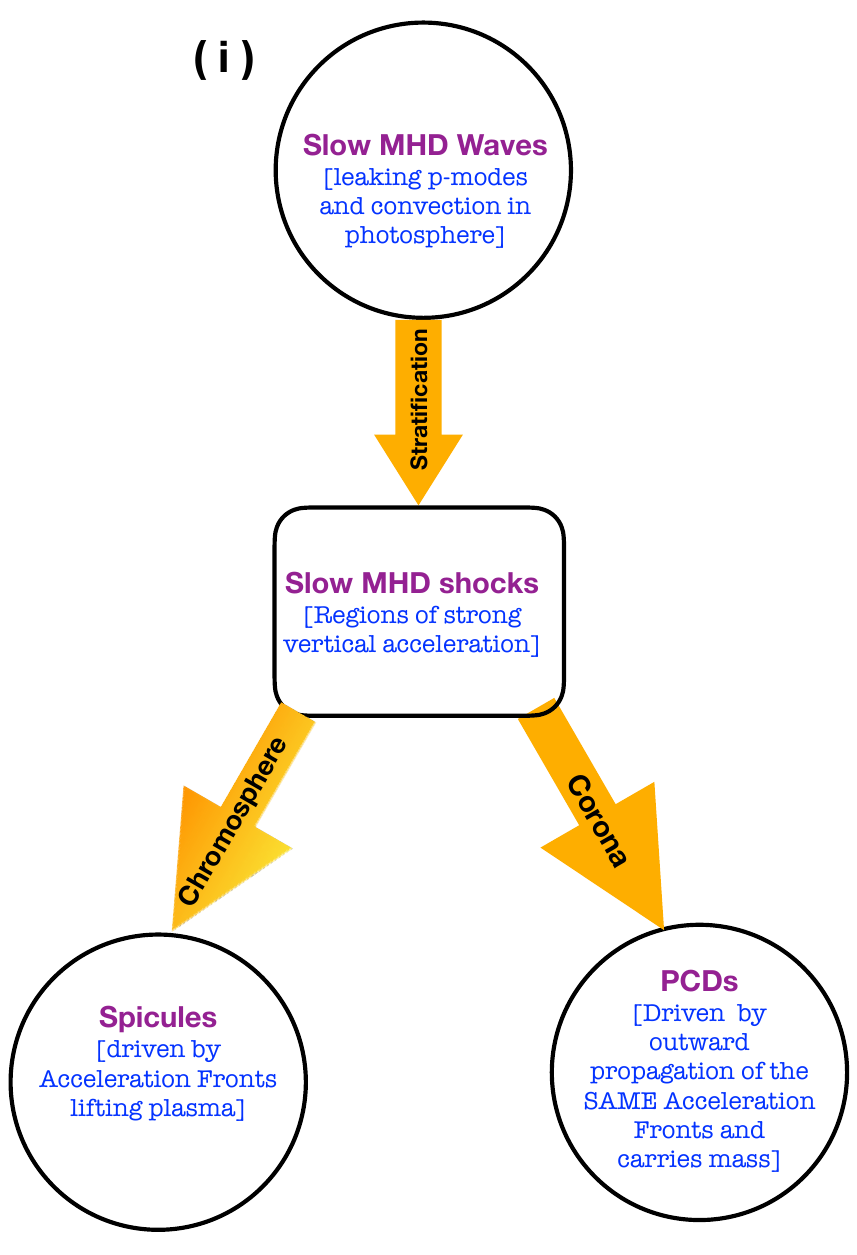}
\caption{(a) Time variation of vertical velocity extracted at 
different heights (see legend) along the vertical slit (of width 
5 grid spacings) placed at $\sim x=6.0$ \,Mm (depicted in Fig \ref{fig:simulation_context}a) 
for the simulation with $B_{\mathrm{imp}}$= 10.0 G. (b)-(d) Time-distance plots along 
the same vertical slit. The synthetic emission for the AIA 171 \AA ~channel in  panel (b) 
shows the coronal intensity enhancements or PCDs, propagating upwards, with the red solid line depicting the sound speed characteristic curve along the slit. In panel (c), we show the 15000 K 
synthetic emission, in which the spicules trace a parabolic trajectory in time. The strong flow convergence or compression regions are overplotted in blue, out of which only the detected slow shocks are indicated in magenta. The red 
horizontal dashed lines show the heights where the 
vertical velocity time series in (a) have been extracted. Panel 
(d) depicts the vertical mass flux along the same vertical 
slit. Panels (e)-(h) are similar to (a)-(d), but for the second 
simulation with $B_{\mathrm{imp}}$=50.0\,G. (i) A schematic to show 
the causality of shock, spicule and PCD connection as deduced from simulations.}
\label{fig:space-time:simulation}
\end{figure*}

To further determine whether these enhanced emission features originate in situ in the corona 
or are imprints of chromospheric structures, e.g., the spicule-driving shocks, we utilize time-
distance diagrams along a vertical slit of 5 grid points width ($\sim$ 0.12 Mm), for both 
runs (panels c and g of Figure~\ref{fig:space-time:simulation}). They reveal 
the 15,000 K emission profile wherein spicules trace a parabolic trajectory in time, 
with the detected slow shocks plotted in magenta. The shocks are seen to move in unison with 
the spicules during their rise phase, demonstrating their role in driving the spicules. It is 
noteworthy that our shock-driven spicule picture is distinct from another existing description 
where the hot tip of spicules induces shocks in the corona~\citep{Mondal22}. 
Furthermore, the strong flow convergence or compression regions, identified in the simulation by setting a 
threshold on the velocity convergence (see Appendix \ref{app:auto_shock}), are overplotted in blue. We note that in both 10 G and 50 G setups, as we move to greater heights, relatively fewer of the strong convergence or compression regions are detected as slow shocks, since they do not exhibit a jump of the flow velocity from superslow to subslow. 

However, many of the compression fronts appear to be continuations of the upward propagating shock waves after they get separated from the spicules and move into the corona.
In other words, often these shock waves (magenta contours) propagating higher up in the 
atmosphere transform to large amplitude non-linear compressive waves (marked in blue iso-
contours). We also find that for the 50 G imposed magnetic field, shock fronts transition to non-
linear compressive waves at relatively lower heights than for the 10 G run (panel g 
compared to panel c of Figure~\ref{fig:space-time:simulation}), likely due to the suppression of convective 
power in the presence of a stronger magnetic field. The Mach numbers calculated for the upstream perpendicular flow at a typical compression front around $z=9.2$ Mm are 1.1 and 0.4 for the 10 G and 50 G cases, respectively (in the rest frame of the front).
The asymmetric or non-linear nature of the waves is confirmed by examining the time variation 
of vertical velocity extracted at different heights along the slit, as indicated by the red 
dashed lines in panels c and g of Figure \ref{fig:space-time:simulation}. The vertical 
velocity profiles, depicted in panels a and e of Figure \ref{fig:space-time:simulation}, 
show steepening into a sawtooth or N-shaped profile as we go up in the atmosphere, signaling the development of a non-linear wave. 
Notably, the sawtooth profile is present at the lower heights where spicules (and slow shocks) can be detected, consistent with 
our observational findings (Figure~\ref{Fig:Merge_fig2}). At upper atmospheric heights, 
the asymmetric or non-linear shape is retained, even though it is not a shock anymore, since 
it does not show a jump in flow velocity from superslow to subslow across it. Furthermore, 
we notice that the magnitude of the vertical velocity is less for the 50 G 
case than for the 10 G one, again indicative of weaker convective power for the higher 
magnetic field. In the time-distance plots (panels b and f of Figure \ref{fig:space-time:simulation}) of AIA 171 \AA ~synthetic emission, we can notice 
the good correspondence (in terms of the slope) of the 
coronal intensity enhancements or PCDs (propagating upwards) with these compression fronts 
(associated 
with strong positive vertical acceleration in our model). The typical characteristic speeds of PCDs are determined from the slopes of the time-distance 
plots, which are 95.8\,km\,s$^{-1}$ and 26.5\,km\,s$^{-1}$ at $z=15$ Mm for 10 G and 50 G imposed 
fields, respectively. The sound speed characteristic curve is 
denoted (in red) for comparison with the PCD propagation speed for both the $B_\mathrm{imp}$ 
cases. For the higher $B_\mathrm{imp}$, we find that 
the propagation speed of the synthetic PCDs falls well below the sound speed in the corona.   

The compression fronts in corona, as 
discussed, are the imprints of spicule-driving slow shocks in the lower atmosphere. Further, the schematic in panel (i) of Fig.~\ref{fig:space-time:simulation} presents a unified picture from our model in which slow MHD waves 
generated by photospheric convection steepen into shocks, lifting chromospheric plasma to 
form spicules while the same acceleration fronts propagate upward, giving rise to PCDs. As these fronts move into the corona, they might 
transition into non-linear MHD waves depending on the ambient magnetic field strength and 
coronal temperature, linking spicules and PCDs as different manifestations of the same shock-driven dynamics across 
atmospheric layers. 

Moreover, these non-linear waves are capable of carrying mass flux outward into the corona, as shown in panels (d) and (h) of 
Figure \ref{fig:space-time:simulation}, depicting time-distance plots of vertical mass flux along the same vertical slit of Figure~\ref{fig:simulation_context}~a.

\begin{figure*}[ht!]
\begin{minipage}[t]{0.6\textwidth}
\vspace{0pt}
\includegraphics[width=\linewidth]{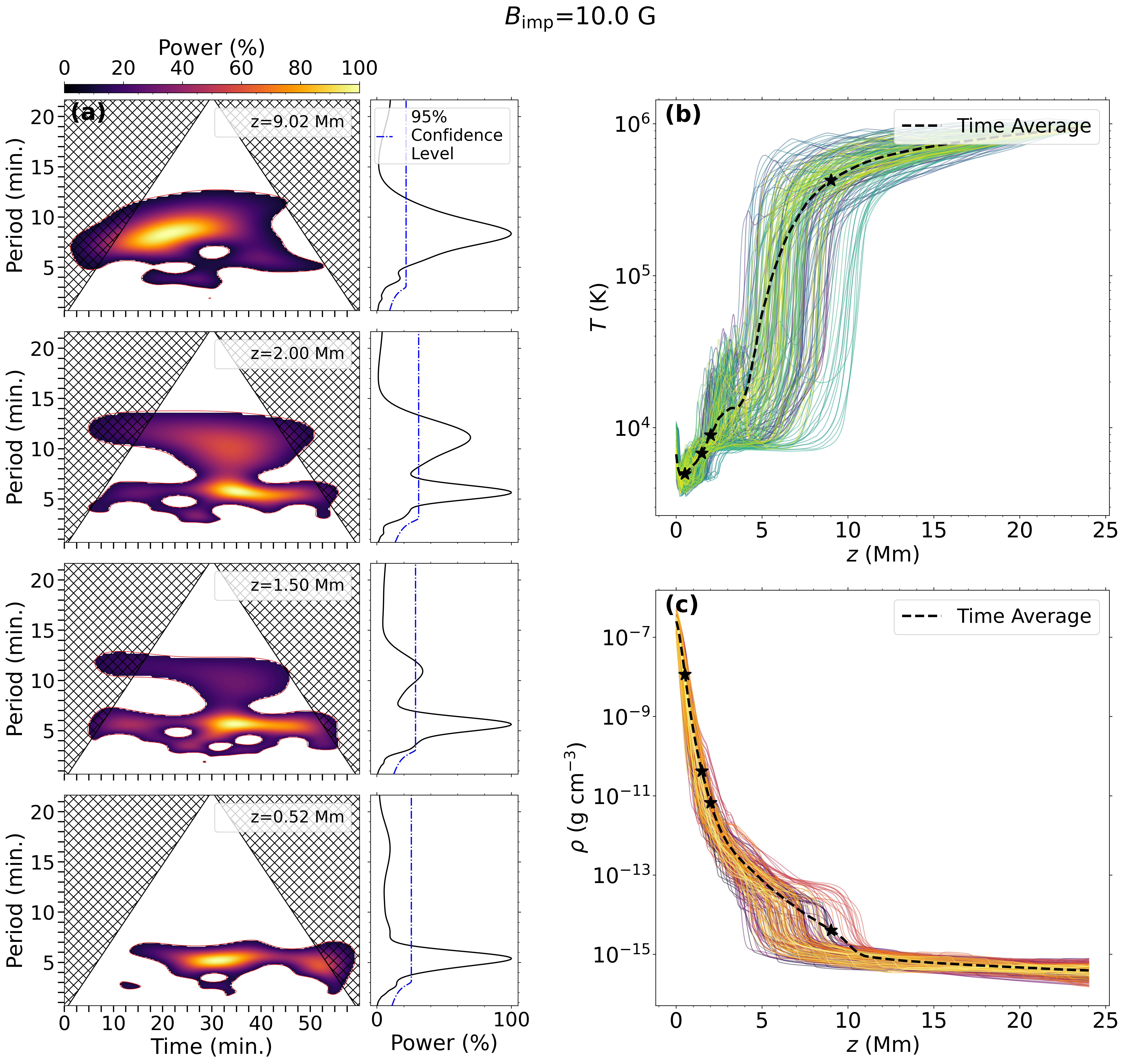}
\end{minipage}
\hfill
\begin{minipage}[t]{0.35\textwidth}
\caption{(a) The wavelet analysis of the vertical velocity time series sampled at 
multiple heights in the $B_\mathrm{imp}=10$\,G simulation run of 
Figure~\ref{fig:space-time:simulation}a. The left panels show the wavelet power 
above 95\% significance level (with areas outside the cone-of-influence marked using a hatched 
pattern), while the corresponding line plots to the right show the global wavelet 
power spectrum. The blue dot-dashed curves on the right panels show the 95\% confidence 
level. (b)--(c) Temperature and density stratification profiles, 
respectively, along the vertical slit of 5 grid points width around $x=6.0$ \,Mm, for the 
run: colored lines correspond to the stratification at different 
times in the 60\,min duration of the synthetic data. The black dashed line shows the time 
averaged variation of temperature or density as a function of $z$. In order to 
assess the effect of the steep stratification on the dominant 
period, the asterix (black) symbols mark the heights above 
the photosphere at which the wavelet analysis is performed.}
\label{fig:wavelet_analysis:simulation}
\end{minipage}
\end{figure*}

The corresponding wavelet 
analysis of the multi-height vertical velocity profiles for 
the 10 G run (panel a of Figure \ref{fig:space-time:simulation}) is shown Figure \ref{fig:wavelet_analysis:simulation}a for different heights. We find the peak or 
the dominant period of the velocity signal is at $\sim$5 
minutes for lower atmospheric heights ($z$=0.5 Mm). The 5-minute period corresponds to the leakage of global solar acoustic oscillations in the lower atmosphere. In the chromospheric region ($z=1.5-2.0$ Mm), a secondary peak, corresponding to $\sim$ 10-minute of periodicity emerge along with the primary 5-minute mode. However, the dominant power completely 
shifts to $\sim$10 minutes of periodicity at the coronal heights (for example, $z$=9 Mm). This transition from lower to the higher dominant period as we move from low chromosphere to corona 
is similar to the observed signals, where the peak power is detected 
at $\sim$ 13 min for the AIA 171 \AA ~channel (highlighted in 
Figure~\ref{Fig:Merge_fig2}). The increase in dominant period behaviour 
might be described in terms of dispersive propagation of waves 
in a highly stratified medium, as shown in panels b and c of Figure \ref{fig:wavelet_analysis:simulation} for temperature and density stratification profiles, 
respectively.
Overall, the 50 G run also shows a similar nature of shift in the dominant period with increasing height in the atmosphere.

\section{Discussion and Conclusions} \label{sec:Discussion} 

In this work, we investigated the origin of a subset of chromospheric spicules that 
appear to be driven by shock waves using concurrent high-resolution observations together 
with two-dimensional radiative MHD simulations. The observations identify these 
events through characteristic sawtooth patterns in $\lambda$--$t$ diagrams at the spicule 
footpoints, a well-known signature of shock formation produced by the nonlinear 
steepening of upward-propagating waves in a stratified atmosphere. 
The key observational result is that the spicules originate from locations exhibiting these chromospheric shock signatures and are followed by upward-
propagating intensity disturbances detected in coronal passbands (Figures 
\ref{Fig:Merge_fig2}, \ref{Fig:Shock_PCD_ex1_space_time} and \ref{Fig6A}--\ref{Fig9A}), suggesting that the 
chromospheric shocks responsible for the launch of the spicules may also generate the coronal 
disturbances observed above them. It is noteworthy that these shock locations do not 
show any clear magnetic flux cancellation in the photosphere, 
as indicated by the magnetograms derived from the PyMilne inversion of the SST Fe I 
6173~\AA\ photospheric line. While the spicules exhibit an average projected velocity of 
$\sim 11$\,km\,s$^{-1}$ in the chromosphere, as inferred from the H$\alpha$ $-0.75$~\AA\ 
filtergrams, the associated PCDs display significantly higher projected velocities of $\sim 
40$\,km\,s$^{-1}$ in the overlying corona.

While these observations reveal a clear temporal and spatial association, the radiative MHD 
simulations provide the physical framework needed to interpret 
these observations. In the simulations, slow MHD waves generated by photospheric 
convection steepen into shocks as they propagate upward through the stratified atmosphere. These 
shocks accelerate chromospheric plasma upward, producing spicular jets whose tips 
coincide with the locations of the detected slow shocks. As the same acceleration fronts continue to propagate into the 
corona, they generate compressive disturbances that appear as intensity enhancements in synthetic EUV emission maps. 
This demonstrates that the same upward-propagating shock fronts that drive the spicules can continue into the corona, and 
naturally produce PCD-like disturbances. \cite{Ni26} have also reported compressive structures associated with spicule upflows and interpret them as slow-mode shocks present at both chromospheric and coronal heights. However, the temporal evolution required to distinguish between upward-propagating shocks and in-situ formation is not explicitly demonstrated. In contrast, our analysis tracks the evolution of disturbances across multiple atmospheric layers, showing that the same upward-propagating shock fronts naturally account for both spicule formation and the associated coronal disturbances.

The simulations capture several key properties inferred from the observations. The shock fronts are 
bands of strong vertical pressure acceleration, which coincide with the tips of the spicules during their 
rise. These upward-propagating shocks or compressive fronts impart impulsive acceleration to plasma parcels as they pass. 
The successively affected plasma parcels evolve ballistically under gravity and magnetic forces, appearing as spicules in chromospheric and transition-region channels. 
The same compressive fronts frequently evolve into large-amplitude non-linear compressive waves as they progress into the corona in the form of Propagating disturbances. Whether these disturbances remain true slow shocks or transition into nonlinear MHD waves depends on the magnetic field strength and the ambient coronal conditions. In either case, the disturbances continue to propagate upward, producing density and temperature perturbations that manifest as intensity variations in coronal channels. 

The simulations also support the observed trend in dominant oscillation period with 
atmospheric height. The chromospheric signal is 
dominated by periods of approximately five minutes, 
consistent with the leakage of photospheric $p$-mode oscillations and MHD wave 
propagation in magnetic flux tubes. At coronal heights the dominant period increases to approximately 10 minutes or 
longer. In the simulations, this shift arises naturally as waves propagate through the strongly 
stratified atmosphere where density decreases rapidly and temperature increases sharply. 
Under these conditions different frequency components propagate at different speeds, producing 
dispersive evolution of the wave packet. In addition, shocks generated at slightly different 
times in the lower atmosphere can merge during upward 
propagation \citep{Chae.et.al.15}, producing longer-period disturbances at higher altitudes.

Because slow MHD waves propagate primarily along magnetic field lines, the coronal response is 
not necessarily located directly above the chromospheric shock 
location. The coronal magnetic field is typically inclined, implying that measurements of 
coronal emission must be taken along the corresponding magnetic field aligned trajectory rather than at 
the vertically projected pixel. 
This effect is illustrated in panels (i)-(j) of Fig.~\ref{Fig:Merge_fig2}, where 
the wavelet power differs when the AIA signal is sampled along the extrapolated magnetic field line compared to the vertically projected location, and in the three-dimensional magnetic field configuration shown in Fig.~\ref{Fig:LFF}. In such cases magnetic-field extrapolation techniques, such as LFFF or related methods, provide a more physically consistent way to trace the field-aligned propagation path and identify the appropriate coronal sampling location.

Another important result concerns the mass flux associated with these 
disturbances. At heights above $\sim15$~Mm the horizontally averaged mass flux in the simulation becomes approximately constant with height, indicating 
that most of the returning chromospheric plasma has already fallen back. The resulting net upward mass 
flux, averaged over five cycles is $\langle\rho u_z\rangle_t \sim (1.1–3.5)\times10^{-11}\,\mathrm{g\,cm^{-2}\,s^{-1}}$. The mass flux is less for the 50 G case than for the 10 G case. To compare the Cartesian simulation with solar-wind observations in spherical geometry, instead of mass flux, we use the conserved quantity of mass-loss rate per unit solid angle, namely, $\langle\rho u_z \rangle r^{2}$, for a radial outflow at $r=R_\odot$. 
This value corresponds to $(6\times10^{10}$–$2\times10^{11})\,\mathrm{g\,s^{-1}\,sr^{-1}}$ which  is comparable to the canonical solar wind value of 
$\dot{M}_\mathrm{WN}\sim7\times10^{10}\,\mathrm{g\,s^{-1}\,sr^{-1}}$, estimated from proton density ($\sim 6$\,cm$^{-3}$) and wind velocity ($\sim650$\,km\,s$^{-1}$) measured at Earth \citep{Withbroe&Noyes.77}. For comparison, the mass flux estimated from both the observed and simulated PCD slopes in $t-z$ images (in Figs.~\ref{Fig:Shock_PCD_ex1_space_time} and ~\ref{fig:space-time:simulation}) with
density inferred from emission measure is $\sim9\times10^{-9}\,\mathrm{g\,cm^{-2}\,s^{-1}}$ and 
corresponds to a mass loss rate per solid angle of $\sim5\times10^{13}\,\mathrm{g\,s^{-1}\,sr^{-1}}$. This is more than two orders of magnitude 
larger than the $\dot{M}_\mathrm{WN}$ value, indicating that this calculation represents a local estimate at an outflow region that does not account for return flows or temporal 
intermittency. Our simulations show that the PCD slope based calculation systematically overestimates the net mass flux, but once the physically relevant averaged quantity is considered, it agrees remarkably well with the solar wind mass-loss rate of $\dot{M}_\mathrm{WN}$.
Here we show that the same upward-propagating acceleration fronts responsible for spicule formation can transport plasma into the corona and generate comparable levels of mass flux. This agreement 
supports the interpretation that the disturbances observed in coronal channels represent a combination of compressive waves and mass motions driven by the same shock dynamics that produce the spicules.

The main advances of this work compared to previous studies can be summarized as follows:

\begin{itemize}

\item By combining high-resolution SST H$\alpha$ observations with SDO/AIA 
coronal imaging, this study traces the evolution of shock-driven dynamics from the low 
chromosphere to the corona. Unlike earlier studies relying mainly on IRIS and AIA 
observations, which primarily probe the upper chromosphere and transition region, our analysis 
provides direct evidence linking chromospheric shock-driven spicules with coronal 
disturbances.

\item We clarify the physical nature of PCDs associated with shock-driven spicules. The 
simulations show that spicule-driving slow shocks can transition into large-amplitude 
compressive waves in the corona, producing intensity disturbances that carry both wave energy and 
mass flux into the upper atmosphere. For this purpose, we rely on the use of an advanced shock detection technique in order to properly assess the role of shocks in the simulations.

\end{itemize}

Taken together, the observations and simulations support a 
unified scenario in which chromospheric shocks generated by convective motions and $p$-mode leakage steepen as they 
propagate upward, drive spicular jets, and continue into the corona as large-amplitude 
compressive disturbances. These disturbances appear observationally as propagating 
coronal disturbances and can transport both mass and energy into the upper solar atmosphere. 
In this framework spicules and PCDs represent different manifestations of the same 
underlying shock-driven dynamics operating across multiple atmospheric layers.

Future work incorporating larger statistical samples covering different regions of magnetic 
geometry, high-resolution three-dimensional simulations, and spectroscopic diagnostics at 
coronal temperatures will be essential to further quantify 
the energy flux carried by these disturbances and to evaluate their role in coronal heating 
and solar wind acceleration. It also remains to be investigated whether other classes of 
spicules, beyond shock-driven events, exhibit a similar connection with propagating disturbances in the corona.

\begin{acknowledgments}
We thank the referee for careful reading of our paper and providing comments that have considerably improved the quality of the presentation. The Swedish 1-m Solar Telescope is operated on the island of La Palma by the Institute for Solar Physics of Stockholm University in the Spanish Observatorio del Roque de los Muchachos of the Instituto de Astrofísica de Canarias. The Swedish 1-m Solar Telescope, SST, is co-funded by the Swedish Research Council as a national research infrastructure (registration number 4.3-2021-00169). We would like to thank SST team for making the data publicly available. SDO is a mission for NASA Living With a Star program. The SDO/HMI data were provided by the Joint Science Operation Centre (JSOC). Further, this work used the DiRAC Data Intensive service (DIaL2) at the University of Leicester and CSD3 at the University of Cambridge (project id: dp261). The DiRAC service at Leicester and Cambridge was funded by BEIS, UKRI and STFC capital funding and STFC operations grants. DiRAC is part of the UKRI Digital Research Infrastructure. PC acknowledges the allocation of computing resources (project ID NAISS 2025/3-69: Astrophysical turbulence and dynamo action) at the PDC Center for High Performance Computing at KTH in Stockholm, funded by the National Academic Infrastructure for Supercomputing in Sweden. The Leo HPC at the Indian Institute of Astrophysics, Bangalore has also been used to perform the numerical simulation and analysis. SD gratefully acknowledges the support provided by the Australian Research Council (ARC) Discovery Project (DP210100709). RE acknowledges the NKFIH OTKA (Hungary, grant No. K142987). RE is also grateful to Science and Technology Facilities Council (STFC, grant No. ST/M000826/1) UK, acknowledges  PIFI (China, grant number No. 2024PVA0043) and the NKFIH (Hungary)  Excellence Grant (grant nr TKP2021-NKTA-64) for enabling this research. This work was also supported by the International Space Science Institute project (ISSI-BJ ID 24-604) on ``Small-scale eruptions in the Sun”.

\end{acknowledgments}

\software{{\sc Pencil Code} \citep{pencilcode2021JOSS}, NumPy \citep{harris.et.al.2020}, SciPy \citep{virtanen2020SciPy-NMeth}, Matplotlib \citep{Hunter:2007}, SunPy \citep{sunpy_community2020}, WaLSA tools \citep{Jafarzadeh2025NRvMP}}

\appendix
\section{Observational Details}\label{app:obs}

\subsection{SST}

The SST dataset comprises imaging spectroscopic observations of the Fe\ \textsc{i}~6173\ \AA, H$\alpha$, and Ca\ \textsc{ii}~8542\ \AA\ spectral lines obtained using the CRisp Imaging SpectroPolarimeter (CRISP; \cite{scharmer.et.al.08}), sampling the photosphere and chromosphere, respectively. The temporal cadence of the three spectral lines was approximately 29.23\,s, with a spatial sampling of $0\farcs0591$ per pixel. The spectral scans consisted of 15, 15, and 17 wavelength positions, respectively, covering ranges of $\pm0.275$\ \AA, $\pm1.75$\ \AA, and $\pm0.7$\ \AA. These observations were processed using the SSTRED reduction pipeline \citep{Lofdahl.et.al.21}. This pipeline employs the Multi-Object Multi-Frame Blind Deconvolution (MOMFBD) technique \citep{Van.Noort.et.al.05} to mitigate the effects of atmospheric seeing and instrumental distortions. By combining numerous short-exposure frames, the method reconstructs images with enhanced spatial resolution and improved image quality. The data reduction sequence further includes dark current removal, flat-field calibration, and accurate co-alignment of the reconstructed frames. These procedures produce calibrated and science-ready CRISP datasets suitable for detailed analysis \citep{delaCruzRodriguez.et.al.15}. 

To infer the photospheric magnetic field, PyMilne Inversion (\cite{delaCruzRodriguez.19}) code has been applied to Fe I 6173 \AA~ observation. Datasets were coaligned by cross-correlating nearly simultaneous wing images dominated by photospheric features. The SST dataset is publicly available through the SST archive\footnote{\url{https://dubshen.astro.su.se/sst_archive/search}}.

\subsection{SDO}

The registered cutouts were obtained using the SDO cutout service, which provides EUV data from the Atmospheric Imaging Assembly (AIA; \cite{Lemen.et.al.12}) and continuum images and magnetograms from the Helioseismic and Magnetic Imager (HMI; \cite{Scherrer.et.al.12}). The AIA UV channels (1600\ \AA, 1700\ \AA) and EUV channels (94–335\ \AA) have nearly identical plate scales of $0\farcs6$\,pixel$^{-1}$, but cadences of 24\,s (UV) and 12\,s (EUV). HMI line-of-sight magnetograms and continuum images have a plate scale of $0\farcs5$\,pixel$^{-1}$ and a cadence of 45\,s. The AIA channels were first coaligned with each other and interpolated to the SST plate scale. Cross-correlation between the nearly simultaneous AIA 1700\ \AA\ and photospheric wing images was then used to establish coordinated alignment with SST observations.

\section{Estimation of Multi-height Doppler Velocities Using the Ca II 8542~\AA\ Spectral Line} \label{sec:Doppler}

\begin{figure}
\includegraphics[width=85mm]{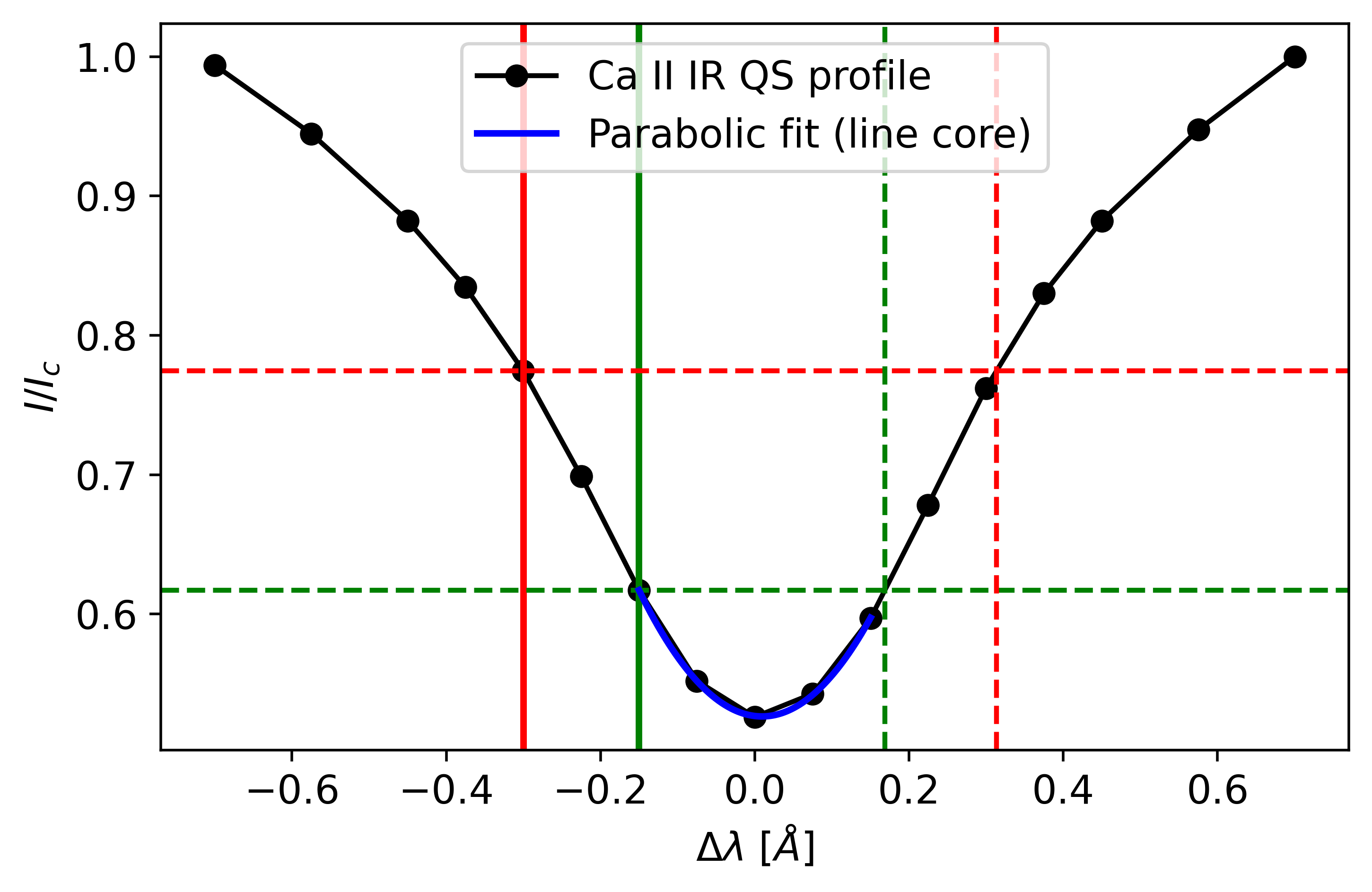}
\caption{The black curve represent the normalized average quiet-Sun Ca~II 8542~\AA. The solid red and green vertical lines indicate the selected wavelength positions in the blue wing ($\Delta\lambda=-0.3$~\AA\ and $\Delta\lambda=-0.15$~\AA) used for the bisector velocity estimation. The dashed horizontal lines mark the corresponding intensity levels, while the dashed vertical lines indicate the interpolated wavelengths in the red wing having the same intensity. The midpoint between the blue and red wavelengths provides the Doppler shift used to estimate the bisector velocities. The blue curve shows the parabolic fit applied around the line core to determine the line-center Doppler velocity.}
\label{fig:avg_CaII}
\end{figure}

The Doppler velocities were estimated at three spectral positions of the Ca~II 8542~\AA\ line corresponding to different formation heights in the chromosphere. These positions are located at $\Delta\lambda=-0.3$~\AA, $\Delta\lambda=-0.15$~\AA, and the line center (LC), representing progressively higher layers of the chromosphere. The selected wavelength positions and the procedure used to estimate the velocities are illustrated on the normalized average quiet-Sun Ca~II 8542~\AA\ spectral profile shown in Figure~\ref{fig:avg_CaII}.

For the wing positions ($-0.3$~\AA\ and $-0.15$~\AA), Doppler velocities were derived using the bisector method. For a given wavelength position in the blue wing, the corresponding intensity level was first determined from the spectral profile. The wavelength in the red wing having the same intensity was then obtained through interpolation of the spectral profile. The Doppler shift was estimated from the midpoint between the blue and red wavelengths,

For the wing positions ($-0.3$~\AA\ and $-0.15$~\AA), Doppler velocities were derived using the bisector method. This approach has been previously employed in earlier work by \cite{Chaurasiya&Bayanna.25}, where the bisector velocities were also compared with other velocity estimation techniques to ensure the robustness of the derived Doppler shifts. In the bisector method, for a given wavelength position in the blue wing, the corresponding intensity level is first determined from the spectral profile. The wavelength in the red wing having the same intensity is then obtained through interpolation of the spectral profile. The Doppler shift is estimated from the midpoint between the blue and red wavelengths.

\[
\Delta \lambda = \frac{\lambda_b + \lambda_r}{2} - \lambda_{\mathrm{ref}},
\]

where $\lambda_b$ and $\lambda_r$ represent the wavelengths in the blue and red wings with identical intensities. The corresponding Doppler velocity was then calculated using

\[
v = \frac{\Delta \lambda}{\lambda_{\mathrm{ref}}} \times c,
\]

where $c$ is the speed of light and $\lambda_{\mathrm{ref}}$ is the reference wavelength.

For the line-center velocity, a parabolic function was fitted to the spectral points around the 
minimum of the Ca~II 8542~\AA\ line profile. The wavelength corresponding to the minimum of 
the fitted parabola was taken as the observed line-center position. The shift of this 
minimum relative to the reference wavelength provides the Doppler shift, which was 
then converted to velocity using the same relation given above.

The reference wavelength $\lambda_{\mathrm{ref}}$ was determined from the average Ca~II 8542~\AA\ profile computed over a quiet-Sun region within the field of view.

\section{Linear Force-Free Magnetic Field Extrapolation}
\label{app:LFF}

To investigate the three-dimensional magnetic field configuration and determine the magnetic field-aligned propagation location in the observations, we performed a LFFF extrapolation using the implementation available in the SolarSoftWare (SSWIDL) package.

The LFFF extrapolation assumes a force-free magnetic field condition,

\begin{equation}
\nabla \times \mathbf{B} = \alpha \mathbf{B},
\end{equation}

where $\alpha$ is a constant force-free parameter that describes the twist of the 
magnetic field. This approximation is generally valid 
in the solar corona, where the plasma-$\beta$ is low and the 
magnetic pressure dominates over the plasma pressure.

As the lower boundary condition, we used the photospheric line-of-sight magnetic field (B$_{\mathrm{LOS}}$) obtained from the PyMilne inversion of the Fe~I 6173~\AA\ spectropolarimetric observations.

The SST magnetic field maps contained regions with missing values (NaNs) near the edges of 
the field of view. To obtain a continuous and physically consistent boundary condition, 
these NaN pixels were replaced with co-aligned HMI line-of-sight magnetograms. The HMI 
magnetograms were spatially co-registered and resampled to match the SST field of view and 
pixel scale. The resulting combined magnetic field map was 
then used as the lower boundary input for the LFFF extrapolation code. This extrapolation provides the three-dimensional magnetic field structure above the photosphere, allowing us to trace magnetic field lines originating from the regions of interest, as shown in Figure~\ref{Fig:LFF}. The extrapolation was performed within a box of size 57~Mm $\times$ 57~Mm $\times$ 8.57~Mm along the $X$, $Y$, and $Z$ directions, respectively. 

These extrapolated magnetic field lines were subsequently used to determine the field-
aligned trajectories along which the wavelet analysis of the AIA 171~\AA\ intensity signal was 
performed. Specifically, the AIA 171~\AA\ signal was taken at a height of approximately 3~Mm 
along the extrapolated magnetic field lines from the lower boundary, corresponding to the 
typical formation height of the lower coronal plasma observed in this channel. This approach 
enables us to investigate the propagation of the shock along the realistic magnetic field 
geometry, rather than assuming purely vertical propagation.

\section{Mass flux carried upward by the PCDs} \label{Sec:mass_flux}

To estimate the mass flux transported upward by the observed PCDs, we performed a differential emission measure (DEM ; \cite{Cheung.et.al.15,Su.et.al.18}) analysis using the optically thin AIA EUV channels. The DEM, in units of $\mathrm{cm^{-5}\,K^{-1}}$, is defined as
\begin{equation}
\mathrm{DEM}(T) = \int n_e^2(T)\, ds,
\end{equation}
where $n_e(T)$ is the electron number density of plasma at temperature $T$ and $s$ is the line-of-sight coordinate.  We integrated the DEM over the temperature range $\log T = 5.7$--$6.1$ to obtain the emission measure,
\begin{equation}
\mathrm{EM} = \int_{T_1}^{T_2} \mathrm{DEM}(T)\, dT.
\end{equation}
Assuming that the emitting plasma is approximately uniform along the line of sight, the electron number density was estimated as
\begin{equation}
n_e = \sqrt{\frac{\mathrm{EM}}{s}},
\end{equation}

where the effective line-of-sight depth $s$ was taken to be three AIA pixels, corresponding to the typical width of the observed coronal loop.
The mass density of the coronal plasma was calculated assuming a fully ionized hydrogen helium plasma with a helium-to-hydrogen number density ratio of $\alpha = n_{\mathrm{He}}/n_{\mathrm{H}} \approx 0.1$ (\cite{Asplund.et.al.09,DelZanna.25}). Under this assumption, the mass density can be written as
\begin{equation}
\rho \approx 1.2\, m_p\, n_e,
\end{equation}
where $m_p$ is the proton mass.

Finally, the mass flux associated with the PCDs was estimated as
\begin{equation}
F_m = \rho\, v,
\end{equation}
where $v$ is the projected plane-of-sky velocity of the PCDs measured from the AIA 171~\AA\ observations. The estimated mass flux values are $9.06 \times 10^{-9}$, $6.69 \times 10^{-9}$, and $1.1 \times 10^{-8}$ g\,cm$^{-2}$\,s$^{-1}$ for the three cases shown in Figure~\ref{Fig:Shock_PCD_ex1_space_time}, yielding an average mass flux of $\sim 9 \times 10^{-9}$ g\,cm$^{-2}$\,s$^{-1}$.

\section{Two Additional Examples of Shock-Driven Spicules and the Associated PCDs}\label{app:A2}

\begin{figure}
\includegraphics[width=32mm]{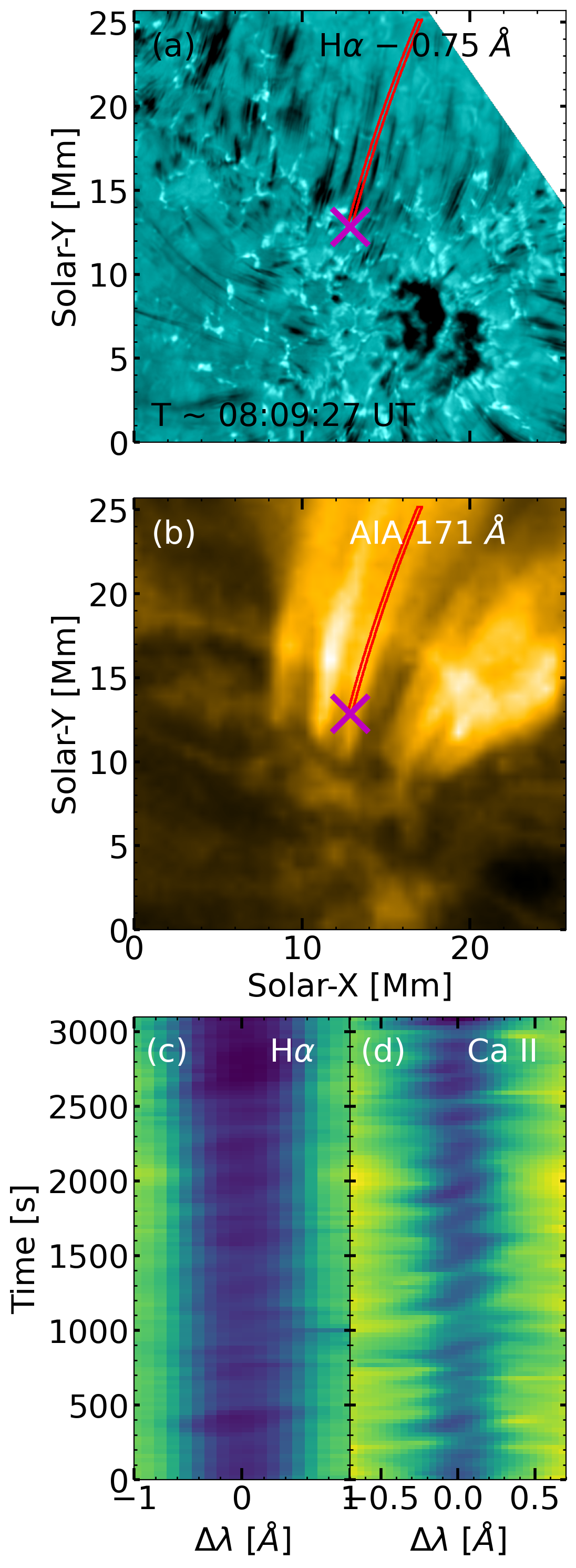}
\includegraphics[width=54mm]{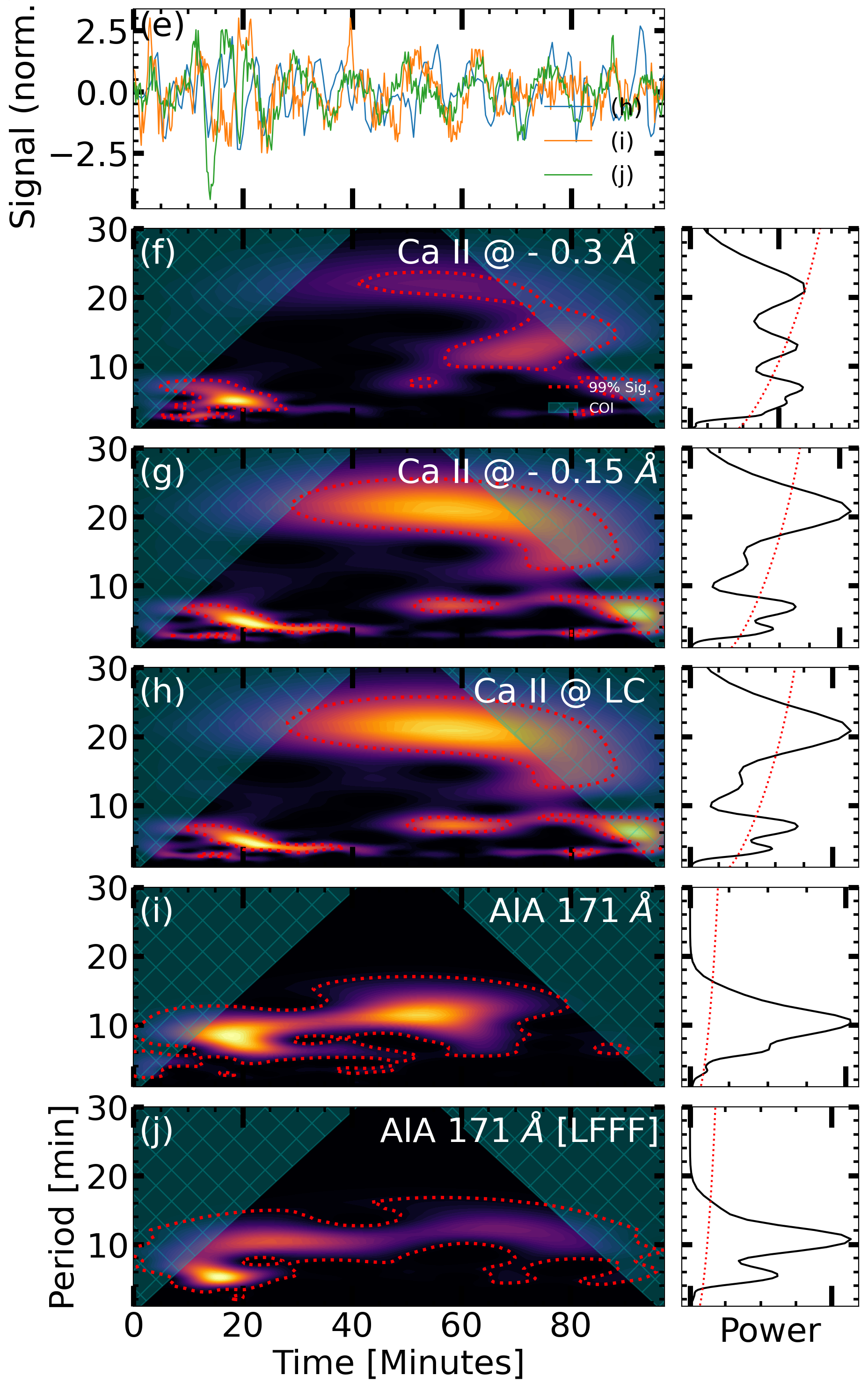}
\caption{Similar to Figure \ref{Fig:Merge_fig2} and also indicated by the green magnetic field line in Figure~\ref{Fig:LFF}.}
\label{Fig6A}
\end{figure}

\begin{figure}
\includegraphics[width=32mm]{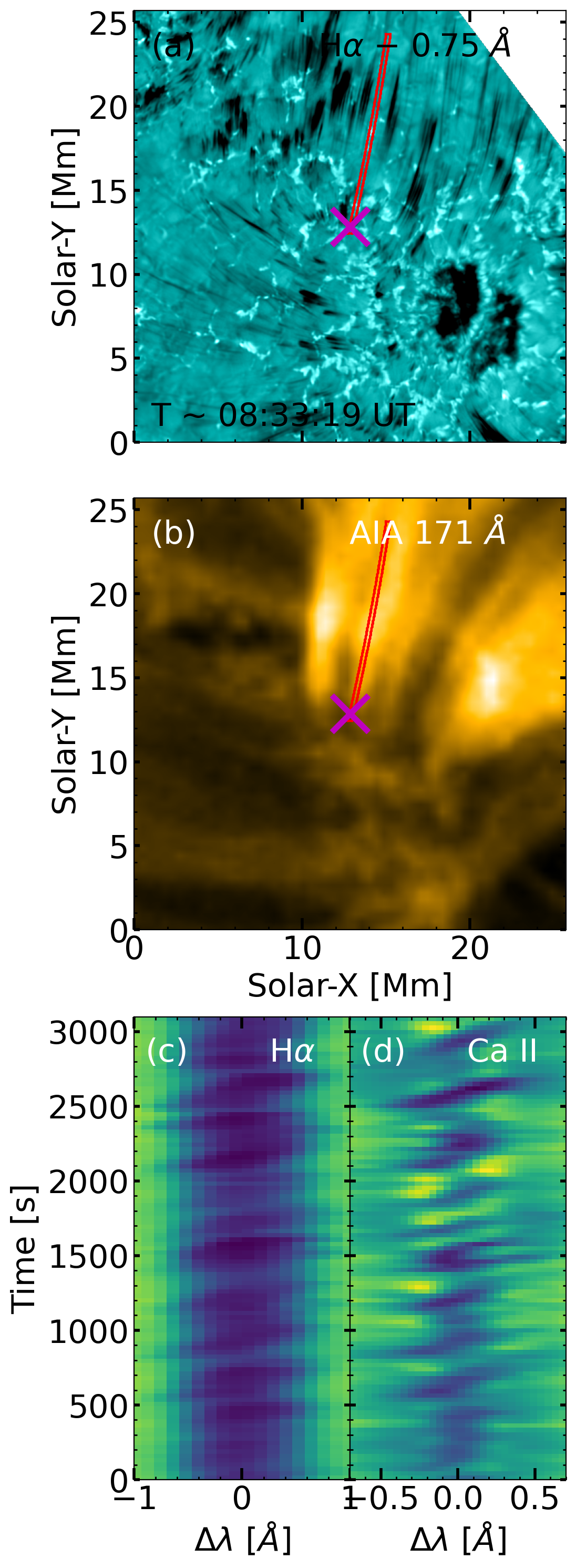}
\includegraphics[width=54mm]{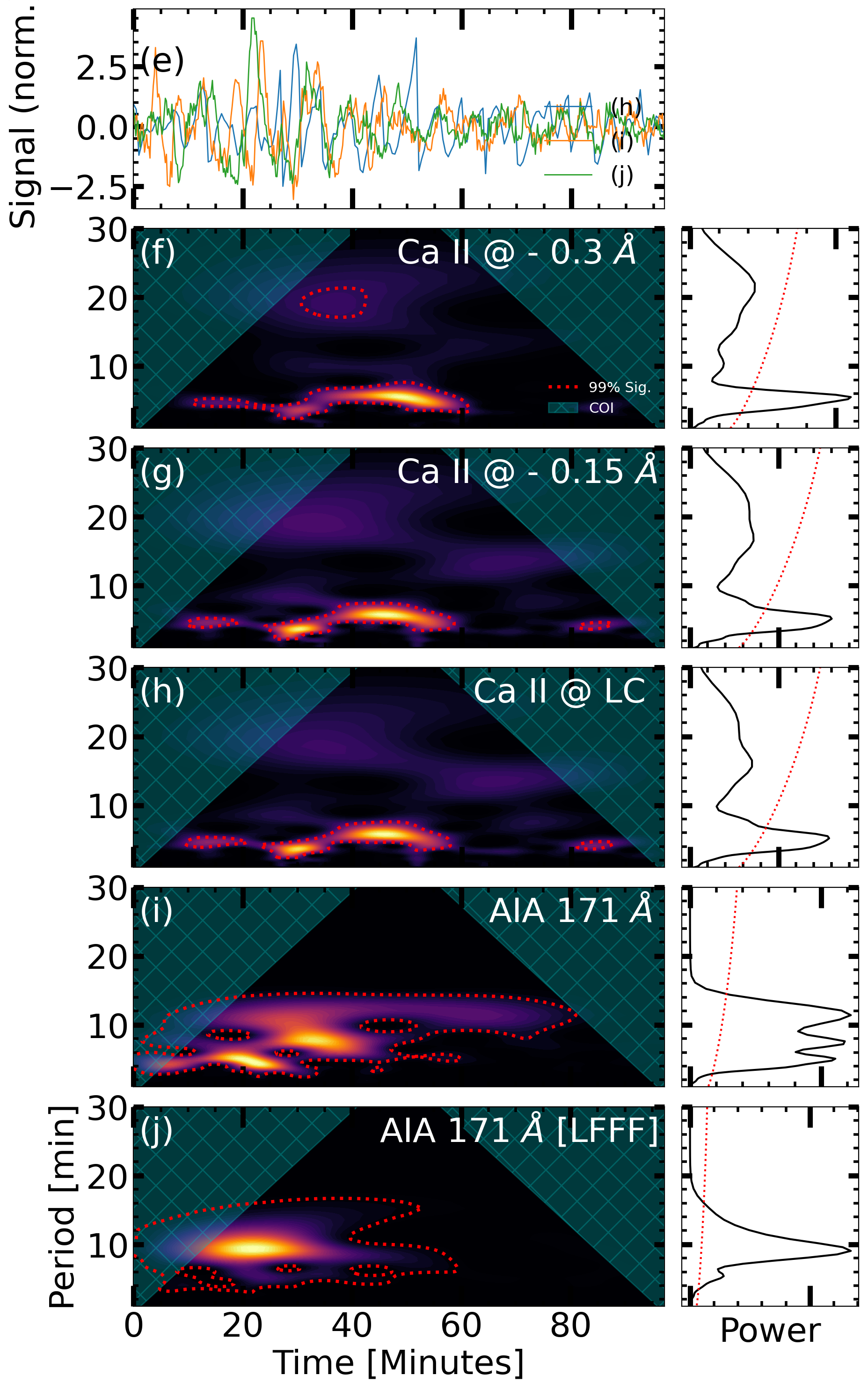}
\caption{Similar to Figure \ref{Fig:Merge_fig2} and also indicated by the blue magnetic field line in Figure~\ref{Fig:LFF}.}
\label{Fig7A}
\end{figure}

\begin{figure}
\includegraphics[width=80mm]{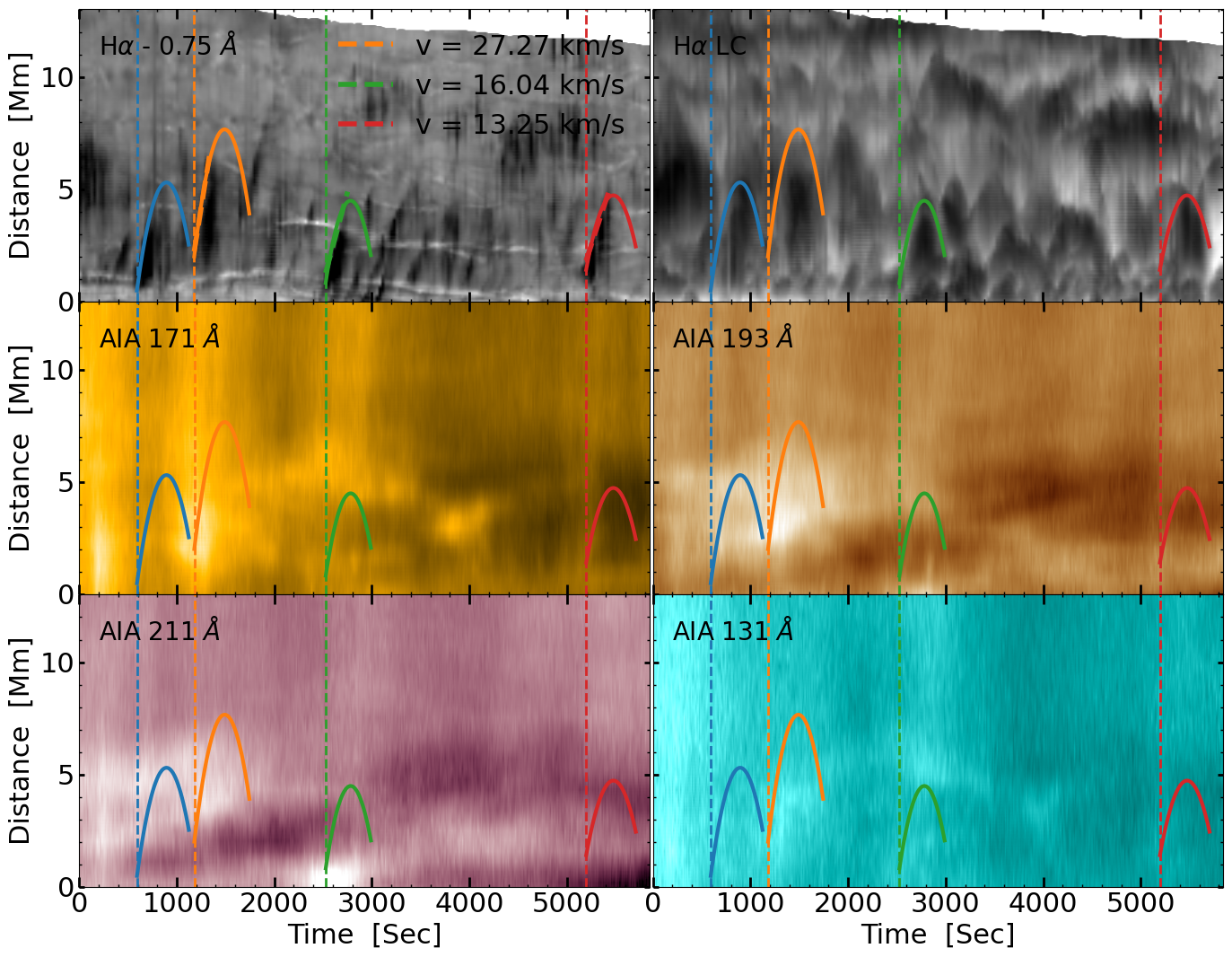}
\caption{Same as Figure \ref{Fig:Shock_PCD_ex1_space_time} but for the slit location shown in Figure \ref{Fig6A}.}
\label{Fig8A}
\end{figure}

\begin{figure}
\includegraphics[width=80mm]{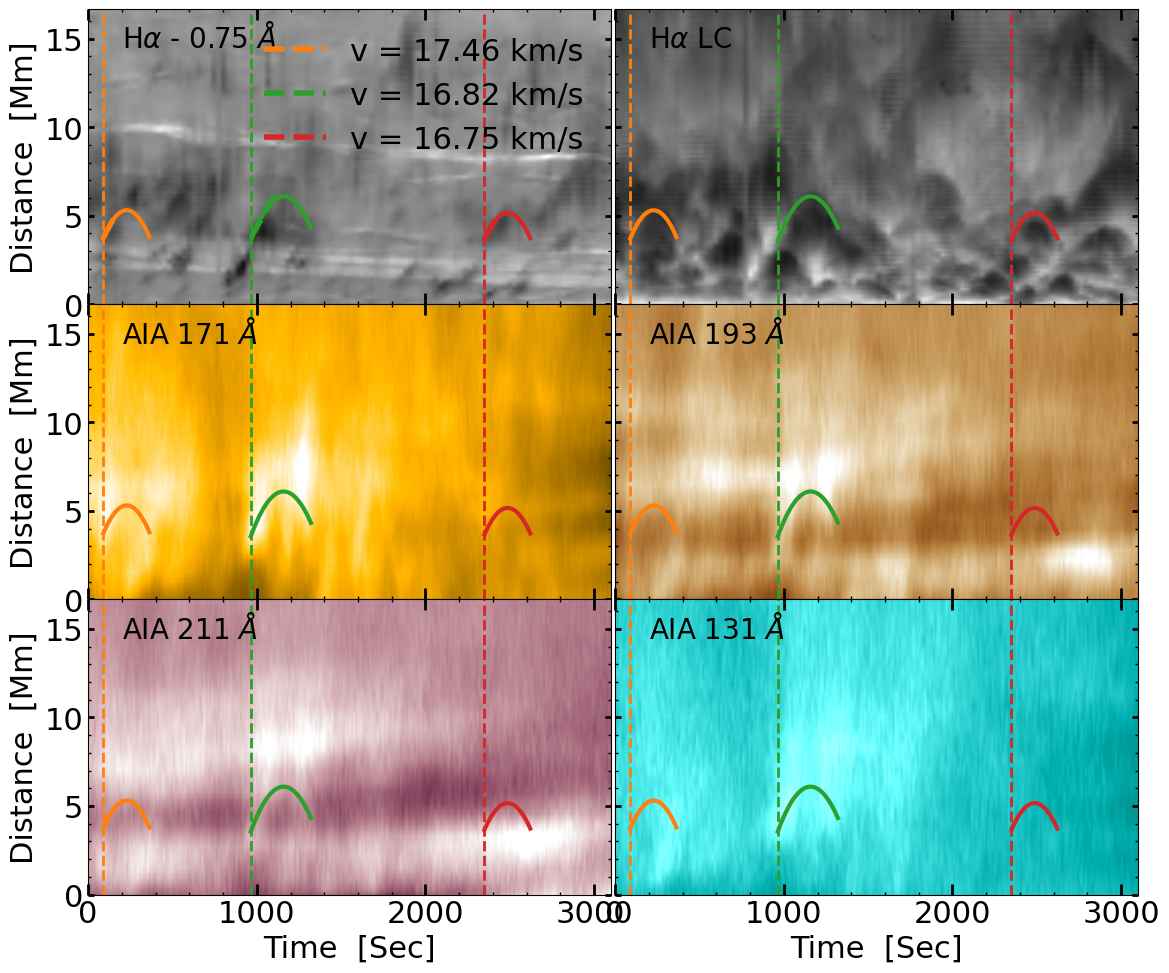}
\caption{Same as Figure \ref{Fig:Shock_PCD_ex1_space_time} but for the slit location shown in Figure \ref{Fig7A}.}
\label{Fig9A}
\end{figure}

\begin{figure*}
\includegraphics[width=60mm]{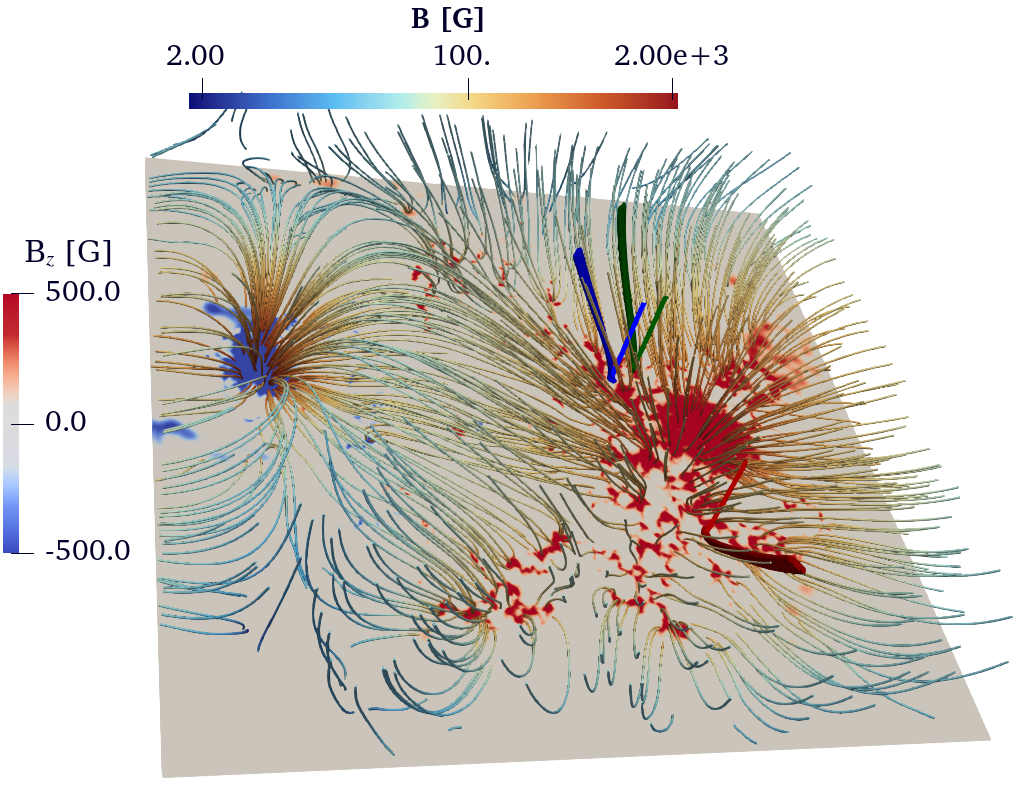}
\includegraphics[width=60mm]{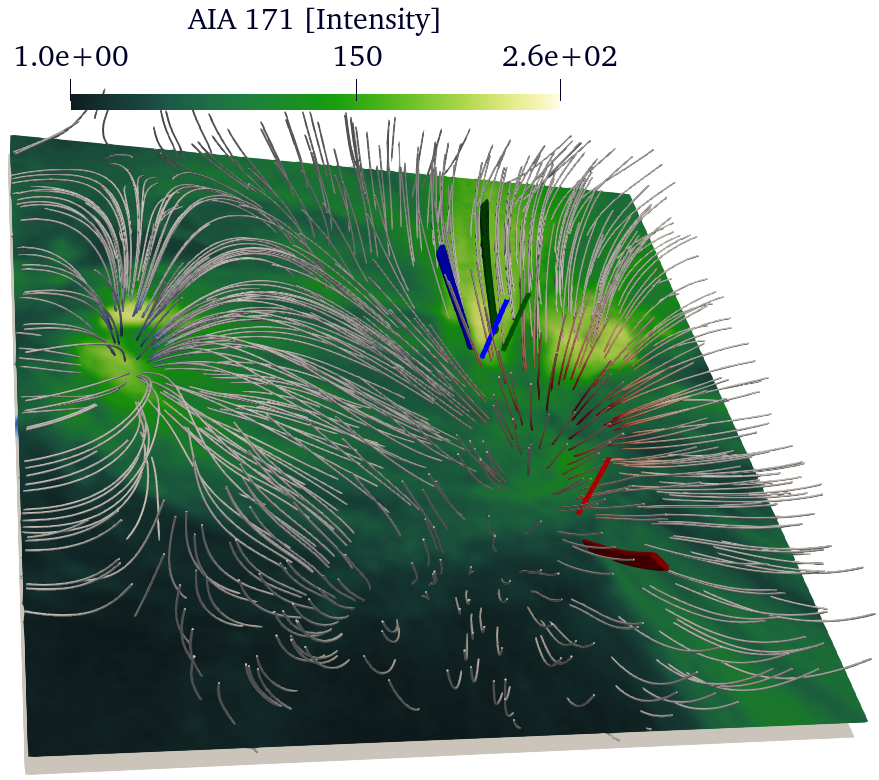}
\includegraphics[width=60mm]{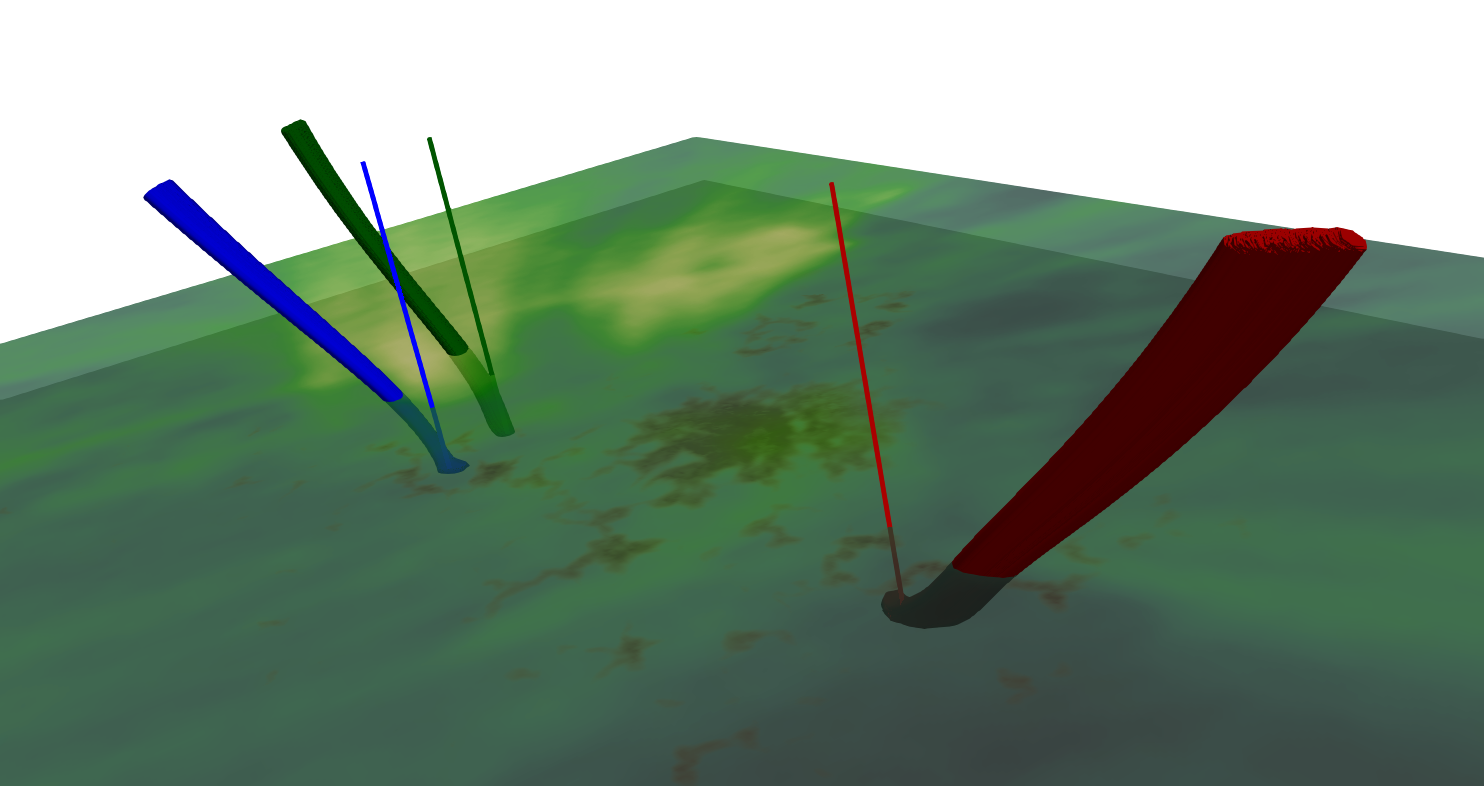}
\caption{Three-dimensional magnetic field structure derived from the LFFF extrapolation, 
with the photospheric magnetic field displayed at the lower boundary. The three thick 
magnetic field lines shown in red, green, and blue correspond to the magnetic field 
orientations at the three shock locations illustrated in Figures~\ref{Fig:Merge_fig2}, \ref{Fig6A}, and \ref{Fig7A}, respectively. The straight 
vertical lines represent the corresponding positions assuming purely vertical propagation. The 
left panel presents the full 3D magnetic field configuration together with the magnetic field 
strength $|\mathbf{B}|$, with the photospheric line-of-sight magnetic field shown at the base. The middle panel displays the AIA 171~\AA\ 
intensity map overlaid on the same field configuration at a height of approximately 3\,Mm 
above the photosphere. The right panel provides a zoomed-in view of the regions containing the analyzed shock wave locations.}
\label{Fig:LFF}
\end{figure*}

Two more examples of shock-driven spicules from observations, presented in the same format as in Figure~\ref{Fig:Merge_fig2}, are shown in Figure~\ref{Fig6A} and Figure~\ref{Fig7A}, while the associated space–time maps illustrating the evolution of the spicules and related PCDs are shown in Figures~\ref{Fig8A} and \ref{Fig9A}.

\begin{figure}[ht!]
\includegraphics[width=0.5\textwidth]{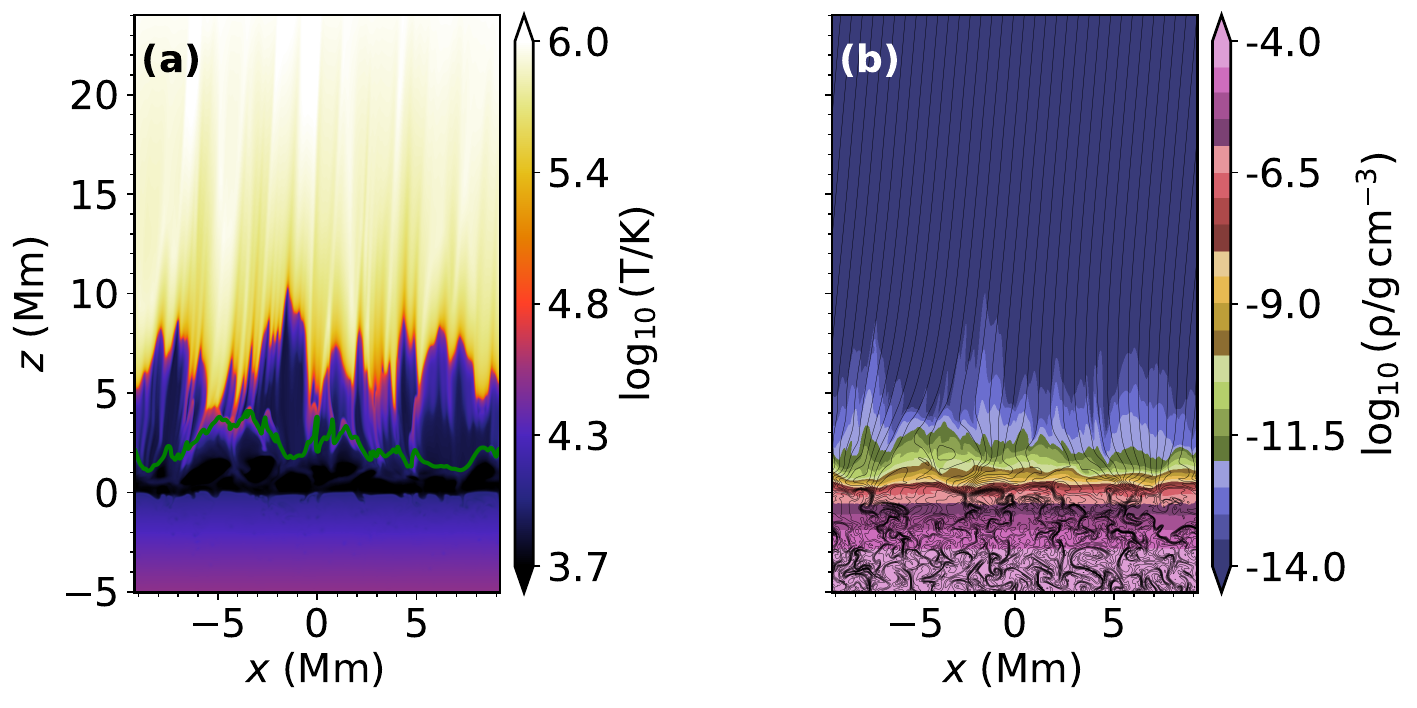}
\caption{(a) Temperature and (b) density distribution at $t=$83.3 minutes from the start of the simulation with an imposed vertical magnetic field $B_{\mathrm{imp}}$= 10.0 G. The plasma-$\beta=1$ surface is over plotted with the green line in (a), while the magnetic field lines are over plotted in black in (b).}
\label{fig:density-temp:simulation}
\end{figure}

\section{Automated Detection of Shocks in Simulations}\label{app:auto_shock}
We developed our own Python version of the IDL code by \cite{Snow_etal_2021} \footnote{The IDL code is publicly available at \href{https://github.com/AstroSnow/Orszag-Tang-Shock-Detection}{https://github.com/AstroSnow/Orszag-Tang-Shock-Detection}} for automated shock identification 
and classification in 2D MHD simulations. The code robustly identifies and classifies all the slow and fast shocks in a given 2D simulation. The first step 
is to identify the candidate shock locations as the ones where the velocity convergence ${\mathbf{-\nabla}\cdot}\uu $ exceeds a threshold value 
appropriate to the simulation run. The unit vectors normal and parallel to the shock at the candidate location are then calculated using the density 
gradient. The locations that do not correspond to local maximum of the normal derivative of density are filtered out to avoid multiple detections due to 
finite shock width. By default, we consider 3 grid points on both sides of the shock candidate location 
along the normal direction for calculations. Next, the shock velocity $v_s$ is estimated from mass 
conservation using the upstream (u) and downstream (d) values of density $\rho$ and perpendicular flow velocity $u_\perp$ as $v_s=[(\rho^\mathrm{d} 
u_\perp^\mathrm{d}-\rho^{\mathrm{u}} u_\perp^{\mathrm{u}})/({\rho^\mathrm{u} -\rho^\mathrm{d}})]$. Then by using the perpendicular flow velocity transformed to the shock frame 
through $v_s$, the code calculates upstream and 
downstream Mach numbers for the perpendicular flow based on the characteristic linear wave 
speeds for slow and fast ideal MHD waves. If we find a transition in the corresponding Mach number 
from values $>1$ to $<1$ across the compression, the candidate is 
identified as a shock location. It can also be classified as a slow or a fast shock.

For both the simulation runs in this study, we take a threshold value of $5\times 10^{-2} \  \mathrm{s}^{-1}$ for velocity convergence, that is reasonably 
low in order to identify all possible shock candidates. The strong flow convergence regions 
plotted in blue in panels c and g of Figure \ref{fig:space-time:simulation} are the regions 
where the convergence exceeds this threshold value. We note that the final shock 
identification is fairly independent of this initial threshold since it is based on 
Mach number calculations. We focus only on the detection of slow shocks in the simulations 
performed here.

\section{Calculation of Synthetic Emission}\label{app:syn_int}
The synthetic intensity was calculated using the formula:
\begin{equation}
\label{eq:syn}
I_\lambda=\int G_\lambda(\rho,T) \rho^2 ds,
\end{equation}
while ignoring the line-of-sight integration (that is, along $ds$) for our 2D simulations. For the AIA 171 \AA ~emission, we used the corresponding temperature response function for $G_\lambda(\rho,T)$. The expression for $G_\lambda(\rho,T)$ for the 15,000 K emission, along with more details, can be found in \cite{Sankalp.srivastava.et.al.25}.

\bibliography{article}{}
\bibliographystyle{aasjournalv7}

\end{document}